\pdfoutput=1

\documentclass[journal,twoside]{IEEEtran}
\usepackage{cite}
\usepackage{graphicx}
\graphicspath{{./} {fig/}}
\usepackage[tight,footnotesize]{subfigure}
\usepackage{amsmath}
\usepackage{algorithmic}
\usepackage{url}
\usepackage{array}

\usepackage{threeparttable}
\usepackage{tabularx}
\usepackage{booktabs}
\usepackage[table,xcdraw]{xcolor}
\usepackage{adjustbox}
\usepackage[utf8]{inputenc}

\usepackage{acronym}
\usepackage{multirow}
\usepackage{algorithm}
\usepackage{algorithmic}

\usepackage{verbatim} %
\begin{document}

\title{A $0.11-0.38$ pJ/cycle Differential Ring Oscillator in $65$ nm CMOS for Robust Neurocomputing}

\author{Xueyong~Zhang,~\IEEEmembership{Student~Member,~IEEE,}
        Jyotibdha~Acharya,~\IEEEmembership{Student~Member,~IEEE,}\\
        and~Arindam~Basu,~\IEEEmembership{Senior Member,~IEEE}
\thanks{Manuscript received January 19, 2020; revised March 26, 2020.}
\thanks{X. Zhang, J. Acharya and A. Basu are with the School of Electrical and Electronic Engineering, Nanyang Technological University, Singapore 639798 (e-mail: xueyong001@e.ntu.edu.sg; arindam.basu@ntu.edu.sg).}
}

\maketitle
\IEEEpeerreviewmaketitle

\begin{abstract}

This paper presents a low-area and low-power consumption CMOS differential current controlled oscillator (CCO) for neuromorphic applications. The oscillation frequency is improved over the conventional one by reducing the number of MOS transistors thus lowering the load capacitor in each stage. The analysis shows that for the same power consumption, the oscillation frequency can be increased about $11\%$ compared with the conventional one without degrading the phase noise. Alternatively, the power consumption can be reduced $15\%$ at the same frequency. The prototype structures are fabricated in a standard $65$ nm CMOS technology and measurements demonstrate that the proposed CCO operates from $0.7-1.2$ V supply with maximum frequencies of $80$ MHz  and energy/cycle ranging from $0.11-0.38$ pJ over the tuning range. Further, system level simulations show that the nonlinearity in current-frequency conversion by the CCO does not affect its use as a neuron in a Deep Neural Network if accounted for during training.

\end{abstract}
 

\section{Introduction}
\label{sec:intro}

\IEEEPARstart{D}{eep} Neural Networks (DNN) are gaining popularity recently in many applications such as face recognition\cite{dnn_face}, speech recognition\cite{dnn_speech}, natural language processing\cite{dnn_nlp} etc due to improved performance compared to other machine learning algorithms. Deploying these networks at the edge in the Internet of Things (IoT) is important for scalability and fast response by reducing data transmission to the cloud\cite{basu_lora,basu_bmi_chapter}. However, due to battery life and processing constraints, it is imperative to have low power, low area implementations of the DNN.

Neuromorphic implementations that utilize analog or physical computing are good in that context\cite{basu_jetcas} and have been known to be energy efficient compared to digital baselines when the required precision is low\cite{sarpeshkar_anavsdig}. Neuro-inspired spiking neural networks (SNN) have also gained popularity due to the promise of sparse activation leading to lower energy dissipation\cite{snn_pfeiffer}. In recent years, time-based computational circuits for DNN/SNN are gaining popularity due to the reduced power supply in scaled CMOS \cite{TDNN_bibhu,TDNN_ASSCC18}. An important building block in these designs is a digital delay cell.
For example, it is used to create an oscillator that can convert analogue current to digital output (rate based neuron) in \cite{elmpuf_tcas1,elm_enyi,elm_patil} or be used as an integrate and fire neuron\cite{TDNN_bibhu} with bio-plausible refractory period and spike frequency adaptation features. The major requirements for an oscillator to function as a neuron are:
\begin{itemize}
    \item Low area requirement since a large number of them are needed per chip.
    \item Low energy/cycle to reduce operational energy.
    \item Since these circuits typically co-exist with a large amount of digital circuits, they have to be robust against power supply noise and other interference.
\end{itemize}
Interestingly, for NN implementations, linearity of the tuning curve of the oscillator-based neuron is not important since learning can be used to correct for it. 

Given these requirements, a CMOS ring oscillator (RO) based structure seems like a good choice due to its low-area and power requirements. To address the third point of robustness, differential ring oscillators should be the preferred topology \cite{pseudo_RO}. In this paper, we propose a new differential delay cell that has reduced number of transistors compared to the conventional one thus reducing energy/cycle due to lower capacitance. {Note this is different from \cite{cicc07} where the authors demonstrate reducing the number of stages.} Given the popularity of rate based SNN converted from pre-trained artificial neural network (ANN) \cite{sengupta2019going,rueckauer2017conversion}, we show results from a fabricated chip in $65$ nm CMOS where the proposed delay cell is used in a differential ring oscillator to convert input analog currents to a frequency or rate. Nevertheless, the proposed delay cell can be used in conventional mixed signal applications as well as differential delay line based neural networks\cite{TDNN_ASSCC18}. Further, other neuromorphic computation using coupled oscillators are also gaining in popularity \cite{arijit_piee,arijit_nature}--the proposed oscillators can be used for such computations as well. 

This paper is organized as follows. Section \ref{sec:conv} gives a review of conventional ring oscillator based CCO (RO-CCO) structures. The proposed structure and analysis are introduced in section \ref{sec:prop}. Section \ref{sec:results} presents measurement results and shows a performance comparison. 
Section \ref{sec:discussion} discusses the neural network simulation with the proposed CCO neuron model, its potential for usage in spiking neural networks as well as the startup circuits. 
Finally, the conclusions are provided in section \ref{sec:conclusion}.

\section{Conventional RO-CCO Review}
\label{sec:conv}
A RO is made from gain stages, or delay stages, in feedback. ROs can be built in any standard CMOS process with a number of delay stages connected in a feedback loop. 


\begin{figure*}[!t]
\centerline
{\includegraphics[width=0.9\textwidth]{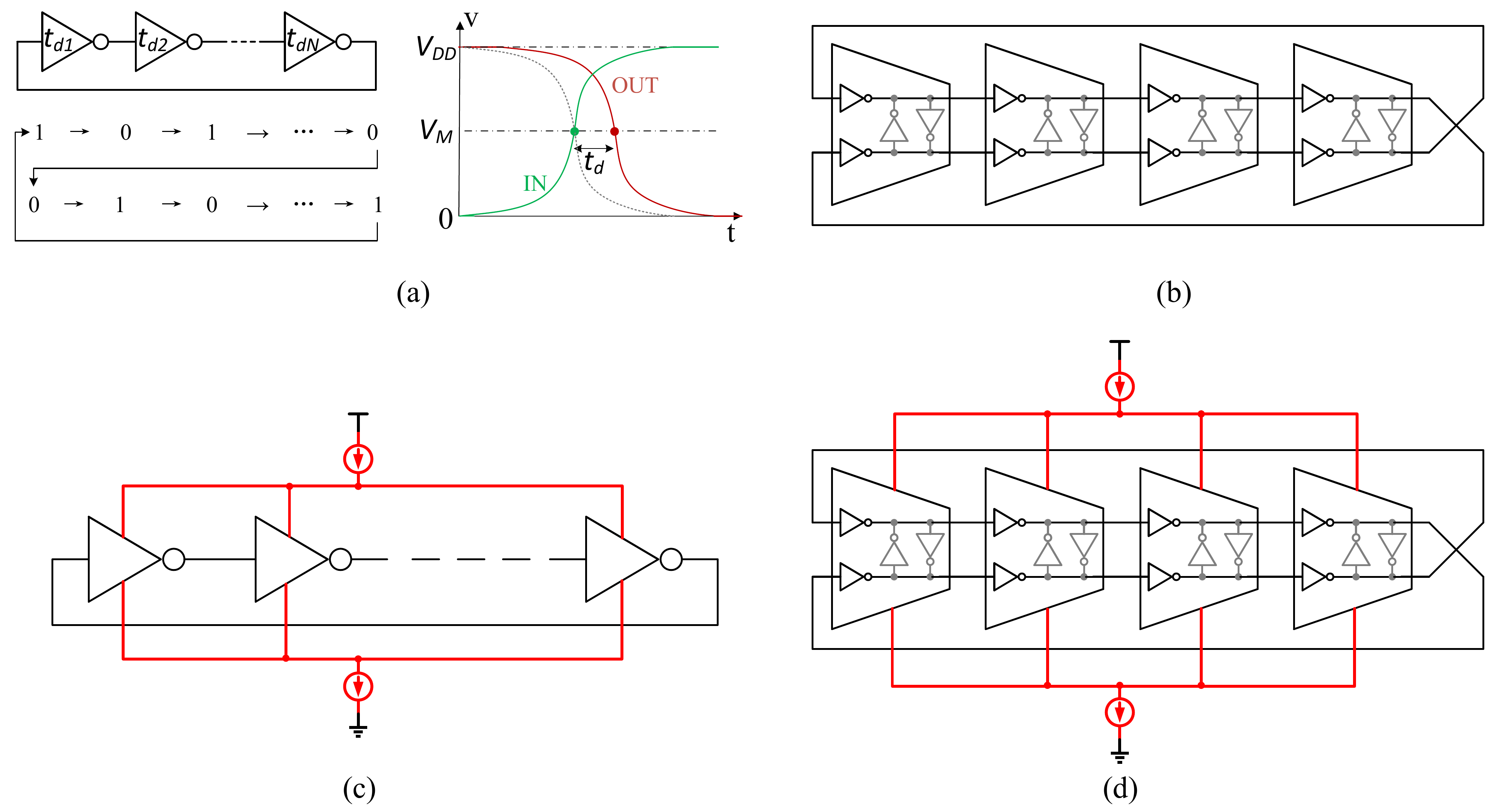}}
\caption{(a) Single-ended RO with odd number of stages and delay definition of each stage. (b) Pseudo differential RO with $N=4$ stages. (c) Current starved single-ended RO. (d) Current starved fully differential RO.}\label{fig:RO_pseuo_diff}
\end{figure*}

A single ended RO is comprised of odd stages of inverters cascaded in self-feedback (Fig. \ref{fig:RO_pseuo_diff}(a)) with capability of rail-to-rail output. 
The most common way to derive an equation for the oscillation frequency of an $N$-stage ($N$ is an odd number) ring oscillator is to assume each stage provides a delay of $t_d$. For a stable oscillator, the signal must go through each of the delay stages twice to provide one period of oscillation, as shown in Fig. \ref{fig:RO_pseuo_diff}(a). 
Therefore, the period is $2Nt_d$, resulting in the oscillation frequency equation

\begin{align}
\label{eq:f}
f = \frac{1}{{2N{t_d}}} = \frac{{{I_{ss}}}}{{2N{C_L}{V_{sw}}}} \propto \frac{{I_{ss}}}{{C_L}}.
\end{align}

Where $I_{ss}$ is the tail current, $C_L$ is the total load capacitance of a single stage, $V_{sw}$ is the amplitude of the voltage swing. Delay of each stage $t_d$ is defined as the time interval of output signal and input signal of the same stage at half VDD, illustrated in Fig. \ref{fig:RO_pseuo_diff}(a). The propagation delay $t_d$ is the most important parameter of this type of oscillator because it directly determines the oscillation frequency, the number of stages and hence determines the number of output phases and the power consumption. Moreover, the stability of delay time $t_d$ also reflect the jitter characteristics in time domain and phase noise in frequency domain.

Eliminating the restriction of odd number of stages in single ended RO, pseudo differential RO is more flexible since both odd and even stages are allowed. A $4$-stage pseudo differential RO \cite{pseudo_RO} is shown in Fig. \ref{fig:RO_pseuo_diff}(b). A single delay stage is composed of four inverters, in which two inverters work as pseudo differential structure, while another two grey inverters constitute cross-coupled regeneration to generate hysteresis phase shift. Differential RO can offer more output phases compared with single ended RO with the same stages. This is a very important criteria in time domain processing circuits especially where multiple phases are necessary for more accuracy operations. 

Although single ended configurations consume less area, they are more susceptible to power supply variations and common-mode noise. In modern machine learning accelerators with analog computing neurons co-existing with digital memory and processing, it is extremely important to ensure good noise rejection properties for the analog blocks. A pseudo differential RO can suppress common mode noise and have a better common mode rejection ratio (CMRR) characteristic \cite{pseudo_RO}. 

To further reduce the sensitivity to power supply variation and noise disturbance, current starved structures, as shown in  Fig. \ref{fig:RO_pseuo_diff}(c), have been introduced \cite{hajimiri1999jitter, dudek2000high}. Since the current source and current sink transistors offer a negative feedback in the bias, the current variation is much less than non current starved structures. Another important advantage is that the current starved structures control the output frequency by controlling the bias current, resulting in the name of current controlled oscillator (CCO).

Finally, combining current starving and differential architecture, current starved fully differential RO (Fig. \ref{fig:RO_pseuo_diff}(d)) has better power supply rejection ratio (PSRR) and CMRR characteristics than other types of RO (shown later in Fig. \ref{fig:line_senstivity}). The power supply noise and substrate noise are isolated by current source and current sink, resulting in a better PSRR. Common mode disturbances such as temperature variations can also be suppressed by the differential structure.



\begin{figure}[!t]
\centerline
{\includegraphics[width=0.45\textwidth]{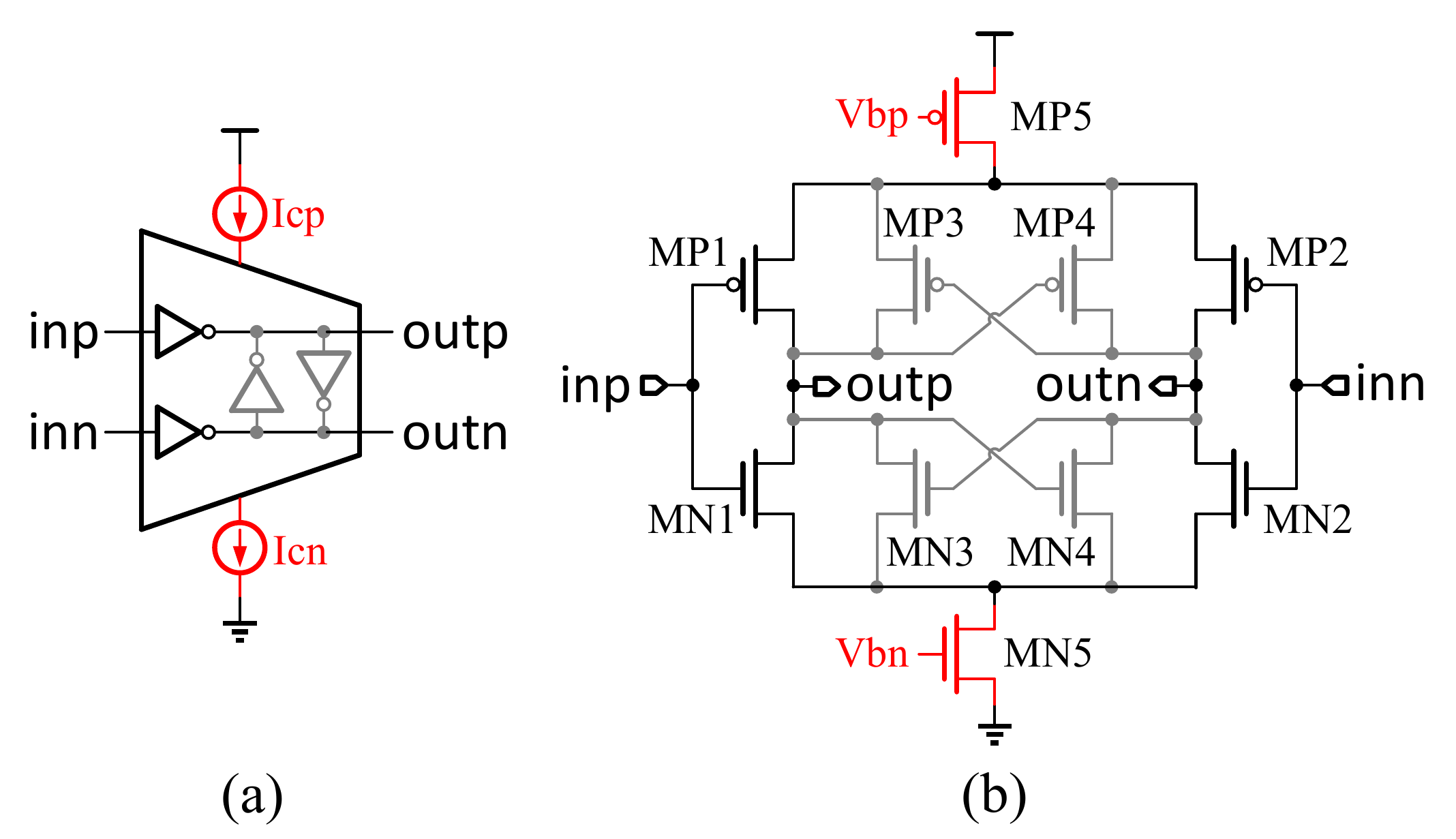}}
\caption{Delay stage of fully differential RO showing the (a) Symbol and (b) Transistor level circuit diagram comprising $8$ transistors excluding the shared current sources on top.}\label{fig:Delay_Stage}
\end{figure}

Although in practice, current starving transistors are shared by all stages, it is easier to understand the operation of an unit delay cell by assigning a current source and a current sink to it. Fig. \ref{fig:Delay_Stage} shows the single delay stage of current starved fully differential RO. MN1, MP1 and MN2, MP2 constitute the differential input and differential output pair. MN3, MP3 and MN4, MP4 constitute the positive feedback latch to offer hysterestic phase shift. 
The controllable current source and current sink, MP5 and MN5, make the delay stage from pseudo differential structure to fully differential. The core stage is composed of a differential input-output path and a hysteresis delay cells (HDC) composed of two cross-coupled inverters based positive feedback latch.

\section{Proposed Structure and Theoretical Analysis}
\label{sec:prop}
As mentioned in Section \ref{sec:intro}, two important metrics for oscillators used for neurocomputing are: (a) high resistance to common-mode noise and (b) low energy/cycle. While the former was addressed through a current starved fully differential architecture as mentioned in the previous section, we address the latter in this section.  However, lowering energy is often associated with an increase in thermal noise induced jitter. We propose a new oscillator structure that is better than conventional structure in terms of energy/cycle while not adversely affecting the jitter performance.

\begin{figure*}[!t]
\centerline
{\includegraphics[width=0.5\textwidth]{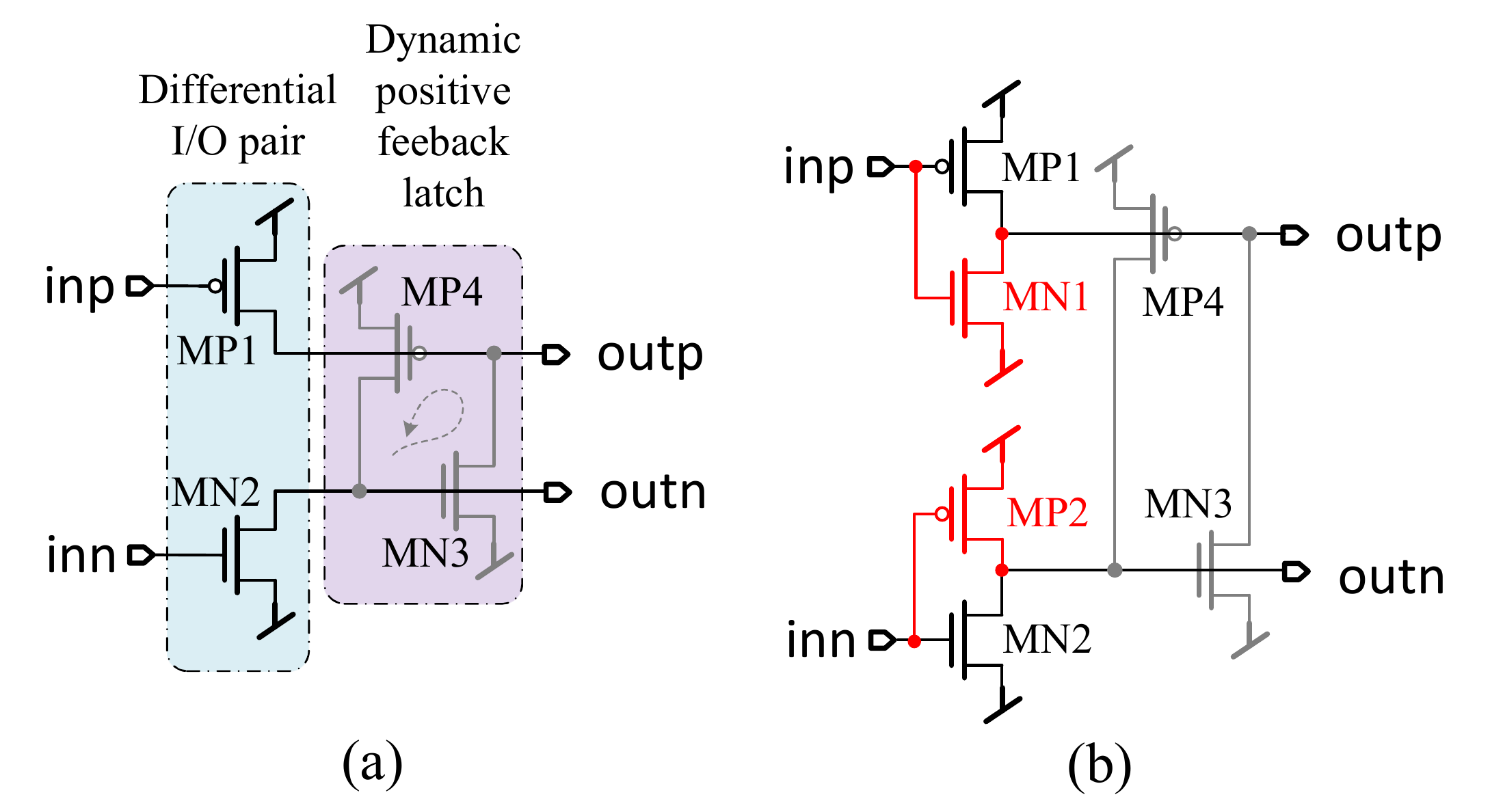}
\includegraphics[width=0.5\textwidth]{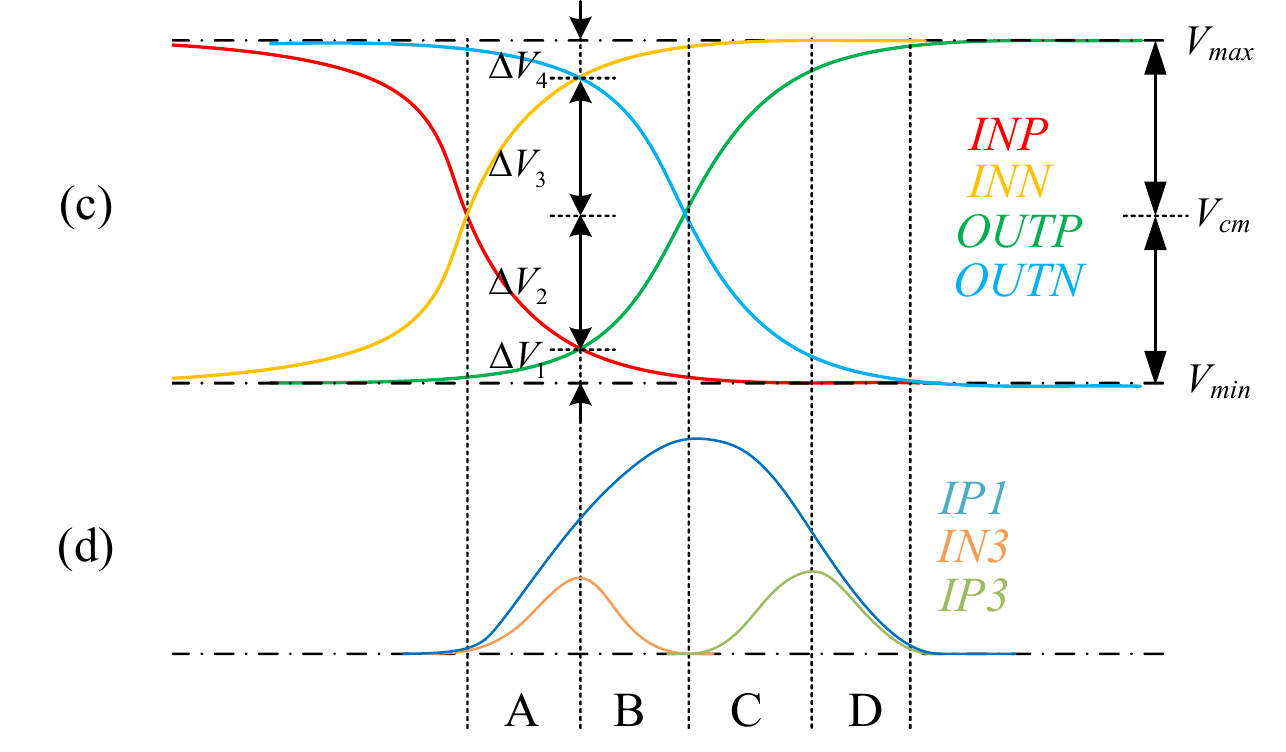}}
\caption{Proposed differential stage of RO: (a) The simplest positive feedback element removing $4$ transistors from the one in Fig. \ref{fig:Delay_Stage}(b). (b) Including $2$ more transistors for the start-up circuit. (Note that current starved transistors are not shown in (a) and (b) for ease of understanding; the sources of PMOS and NMOS are connected to the soft rail voltages $V_{max}$ and $V_{min}$ respectively.) (c) Voltage waveform of the different nodes at the input and output of a delay stage. (d) Charging and discharging currents corresponding to the charging node $outp$ ($IP3$ is the current of MP3 in Fig. \ref{fig:Delay_Stage} (b).)}\label{fig:prop_RO}
\end{figure*}

\subsection{Proposed Oscillator}
\begin{figure}[!t]
\centerline
{\includegraphics[width=0.45\textwidth]{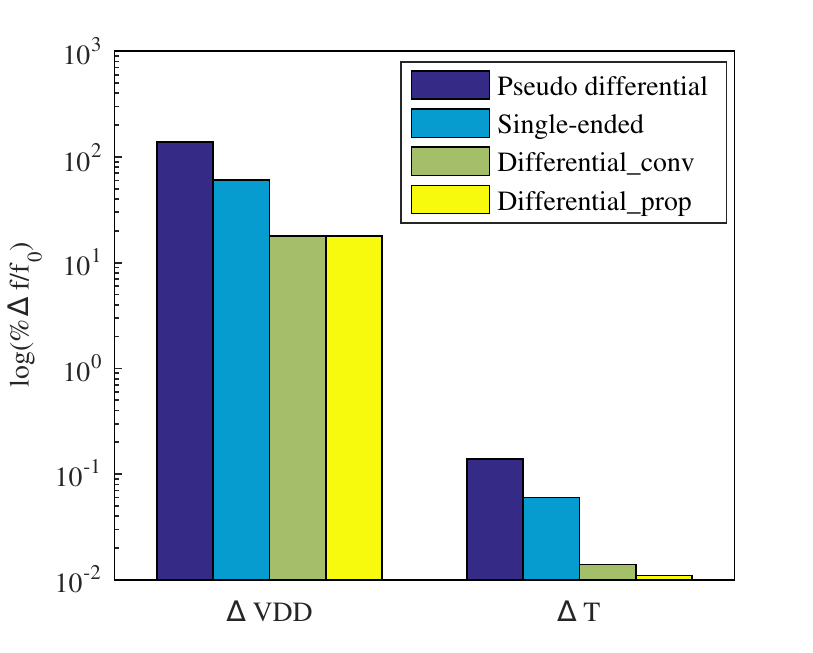}}
\caption{Comparison of the sensitivity of four different types of CCO topology with power supply and temperature: pseudo differntial, current starved single-ended, conventional and proposed current starved differential CCO. The proposed and conventional current starved differential designs have similar sensitivity that is much less than pseudo-differential or current starved single ended architectures.}\label{fig:line_senstivity}
\end{figure}

From the analysis of conventional ring oscillators in section \ref{sec:conv} and simulations shown later in this section (Fig. \ref{fig:line_senstivity}), we know that current starved fully differential structure has both better PSRR and CMRR. 
To offer differential inputs, outputs and a positive feedback hysteresis, each stage of the conventional differential RO uses four inverters (i.e. $8$ transistors) except for the current starved transistors (Fig. \ref{fig:Delay_Stage}). Our novelty stems from the observation that only four transistors seem necessary to realize oscillation as shown in Fig. \ref{fig:prop_RO}(a). MP1 and MN2 are the pull-up and pull-down differential I/O pair while MN3 and MP4 serve as their loads respectively. At the same time, MN3 and MP4 constitute the simplest dynamic positive feedback latch. Another way to interpret this proposed structure is as follows: MP1 and MN3 constitute a dynamic inverter since $outn$ and $inp$ are almost the same (with a certain phase shift), similarly, MP4 and MN2 constitute another dynamic inverter since $outp$ and $inn$ are almost the same (with a certain phase shift). Likewise, the two dynamic inverters compose the regeneration cross pair to generate positive feedback and offer extra hysteresis phase shift. Every transistor acts as an active device for itself and a load device for another transistor. The push-pull nature and transistor reuse make the cell compact and efficient. 

However, one problem is that $inp$ and $inn$ are strictly inverse only when oscillations have been sustained, while $outn$ and $inn$ are inverse with a phase delay of $180^\circ/N$. If for some reason, the initial condition of $inp$ and $inn$ (Fig. \ref{fig:prop_RO} (a)) are same at say $0$ V, then $outp$ is guaranteed to be pulled up to $V_{dd}$ but the condition of $outn$ stays at initial condition since both MN2 and MP4 are OFF. If initial condition of $outn$ was $V_{dd}$, then the outputs of this stage are both at high voltage and the oscillator gets locked in this state and cannot start oscillating. So a start-up circuit is necessary to maintain robust oscillations. 
Only a NMOS or PMOS or both (Fig. \ref{fig:prop_RO}(b)) can eliminate the possible stable state and work as start-up circuit. In the above example, when both $inn$ and $inp$ are $0$ V, now with added MP2, it will force $outn$ to $V_{dd}$ forcing the outputs of this stage to be opposite in polarity thus breaking the lock state earlier. From SPICE simulations, the structure with start-up of both NMOS and PMOS as shown as Fig. \ref{fig:prop_RO}(b), gives the best frequency performance (see Section \ref{sec:discussion}) and is adopted in the rest of the paper. 

\subsection{Frequency and Energy Dissipation}
The novelty in our design of reducing the number of transistors should lead to an increased oscillation frequency per unit current. However, the charging currents also change and hence the combined effect is not obvious. Intuitively, the proposed delay stage offers less charging and discharging current by removing MP3 and MN4, which cuts down on power consumption. To precisely compare the charging/discharging current and the effect on the output frequency, we analyze the load capacitance and charging current model of a single stage in more details to present a theoretical prediction.



First, the load capacitor of each stage for the conventional structure (Fig. \ref{fig:Delay_Stage}) is given by: 
\begin{align}
\label{eq:c_conv}
{C_{L\_conv}} &= {C_{gdP1}} + {C_{gdN1}} + {C_{gdP3}} + {C_{gdN3}} \nonumber \\
              &\qquad {} + {C_{gP4}} + {C_{gN4}} + 2{C_{g\_NextStage}} \nonumber \\
              &= 4{C_{gd}} + 4{C_g} \nonumber  \\
              &= 8{C_{gd}} + 4{C_{gs}}.
\end{align}
In comparison, for the proposed structure (Fig. \ref{fig:prop_RO}(b)), it reduces to
\begin{align}
\label{eq:c_prop}
{C_{L\_prop}} &= {C_{gdP1}} + {C_{gdN1}} + {C_{gdN3}} \nonumber \\
              &\qquad {} + {C_{gP4}} + 2{C_{g\_NextStage}} \nonumber \\
              &= 3{C_{gd}} + 3{C_g} \nonumber  \\
              &= 6{C_{gd}} + 3{C_{gs}}.
\end{align}
Thus, ${C_{L\_prop}}=3/4{C_{L\_conv}}$. 
It is therefore reasonable to expect that the load capacitor decreases by $25\%$ since the number of transistors of each stage decreases by the same amount.

Next, we analyze the charging process of the node of $outp$ in Fig. \ref{fig:prop_RO}(b), while the discharging process is same as the node of $outn$. Fig. \ref{fig:prop_RO}(c) illustrates the charging period, which is divided into four phases (A, B, C and D) according to the operation of the transistors. $V_{max}$ and $V_{min}$ are the swing range of the oscillation, which is lower than rail-to-rail because of the current starved structure. In our design, $V_{max}\approx 900$ mV, $V_{min}\approx 300$ mV, and $V_{cm}\approx 600$ mV. For simplicity, we assume that $V_{TN}=|{V_{TP}}|=V_{T}$, $\Delta V_{1}=\Delta V_{4}=50$ mV and $\Delta V_{2}=\Delta V_{3}=250$ mV such that $\sum_{i=1}^{4}\Delta V_i=V_{max}-V_{min}=600$ mV.


\begin{table*}[htbp]
\centering
\caption{Comparison of charging and discharging current of each contributory transistors for a charging node in four phases, where $\beta=\mu_{0} C_{ox}W/L$. $\beta$ of NMOS and PMOS are the same assuming proper sizing to nominally maximize noise margin. Region 0, 1 and 2 refer to cut-off, linear and saturation regimes of operation of the MOSFET.}
\label{tab:i_comparison}
\begin{tabular}{llll}
\toprule[2pt]
\textbf{Phase}     &\textbf{MP1}     &\textbf{MP3}      &\textbf{MN3} \\ 
\midrule[1pt]
\multirow{7}*{A}   & region=2        & region=0	        & region=1 \\
                   & $v_{sg} \uparrow : V_{max}-V_{cm} \rightarrow $  &  & $v_{gs} \downarrow : V_{max} - V_{min} \rightarrow$\\  
                   & \qquad \quad $V_{max}-V_{min}-50mV $  &  & \qquad \quad $V_{max}-V_{min}-50mV $\\
                   & $v_{sd} \downarrow : V_{max}-V_{min} \rightarrow $  &  & $v_{ds} \uparrow : 0 \rightarrow$\\
                   & \qquad \quad $V_{max}-V_{min}-50mV $  &  & \qquad \quad $50mV $\\
                   & $I=1/2\times\beta(v_{sg}-V_{T})^\alpha \uparrow $  &  & $I=\beta(v_{gs}-V_{T})^{\alpha/2}v_{ds} \uparrow$\\
                   & $\overline{I} \approx 0.02\beta$  &  & $\overline{I} \approx 0.01\beta$\\
\midrule[1pt]
\multirow{7}*{B}   & region=2        & region=0	        & region=$1 \rightarrow 3$ \\
                   & $v_{sg} \uparrow : V_{max}-V_{cm}-50mV \rightarrow $  &  & $v_{gs} \downarrow : V_{max} - V_{min} -50mV\rightarrow$\\  
                   & \qquad \quad $V_{max}-V_{min} $  &  & \qquad \quad $V_{max}-V_{cm}$\\
                   & $v_{sd} \downarrow : V_{max}-V_{min}-50mV \rightarrow $  &  & $v_{ds} \uparrow : 50mV \rightarrow$\\
                   & \qquad \quad $V_{max}-V_{min}-V_{cm}$  &  & \qquad \quad $V_{min}-V_{cm}$\\
                   & $I=1/2\times\beta(v_{sg}-V_{T})^\alpha \uparrow $  &  & $I=\beta(v_{gs}-V_{T})^{\alpha/2}v_{ds} \uparrow$\\
                   & $\overline{I} \approx 0.07\beta$  &  & $\overline{I} \approx 0.01\beta$\\ 
\midrule[1pt]
\multirow{7}*{C}   & region=$2 \rightarrow 1$        & region=2	        & region=0 \\
                   & $v_{sg}(max) : V_{max}-V_{min} $  & $v_{sg} \uparrow : V_{max} - V_{cm}\rightarrow$  &\\  
                   & \qquad \quad  & \qquad \quad $V_{max}-V_{min}-50mV$  &\\
                   & $v_{sd} \downarrow : V_{max}-V_{cm} \rightarrow $  & $v_{ds} \uparrow : V_{max}-V_{cm} \rightarrow$  &\\
                   & \qquad \quad $50mV$  & \qquad \quad $50mV$  &\\
                   & $I=1/2\times\beta(v_{sg}-V_{T})^\alpha \uparrow $  & $I=1/2\times\beta(v_{sg}-V_{T})^\alpha \uparrow$  &\\
                   & $\overline{I} \approx 0.07\beta$  & $\overline{I} \approx 0.02\beta$  &\\ 
\midrule[1pt]
\multirow{7}*{D}   & region=1        & region=1	        & region=0 \\
                   & $v_{sg}(max) : V_{max}-V_{min} $  & $v_{sg} \uparrow : V_{max} - V_{cm}\rightarrow$  &\\  
                   & \qquad \quad  & \qquad \quad $V_{max}-V_{min}-50mV$  &\\
                   & $v_{sd} \downarrow : V_{max}-V_{cm} \rightarrow $  & $v_{ds} \uparrow : V_{max}-V_{cm} \rightarrow$  &\\
                   & \qquad \quad $50mV$  & \qquad \quad $50mV$  &\\
                   & $I=\beta(v_{sg}-V_{T})^{\alpha/2}v_{sd} \uparrow $  & $I=\beta(v_{sg}-V_{T})^{\alpha/2}v_{sd} \uparrow$  &\\
                   & $\overline{I} \approx 0.02\beta$  & $\overline{I} \approx 0.02\beta$  &\\ 

\bottomrule[2pt]
\end{tabular}
\end{table*}

\begin{figure}[!t]
\centerline
{\includegraphics[width=0.45\textwidth]{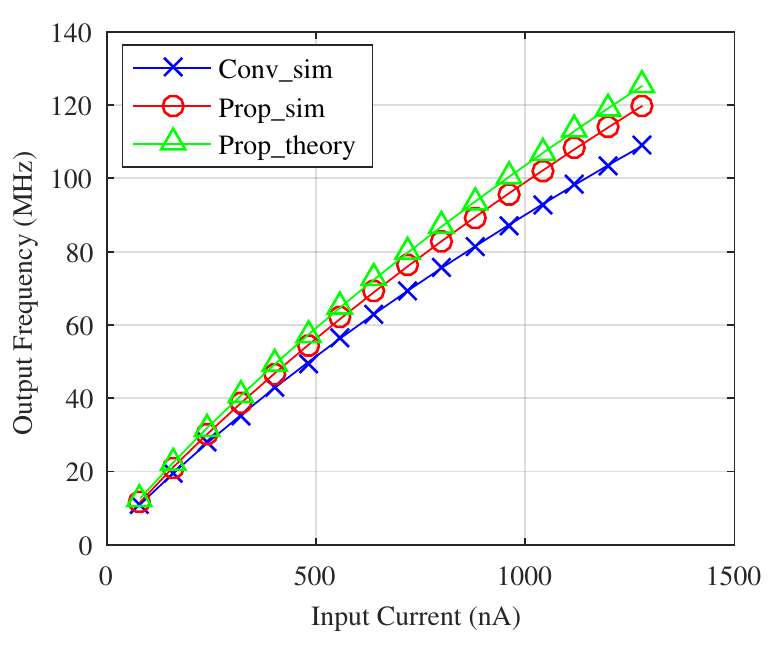}}
\caption{Simulation results of oscillation frequency of both conventional and proposed $4$ stage fully differential RO, along with the theoretical prediction for the proposed structure. As expected, the proposed RO has higher frequency for the same current.}\label{fig:i-f_sim}
\end{figure}

\begin{align}
\label{eq:td_conv}
t_{d\_conv} = \frac{C_{L\_conv} \Delta V_1}{I_{1\_conv}} + \frac{C_{L\_conv} \Delta V_2}{I_{2\_conv}} \nonumber\\
            + \frac{C_{L\_conv} \Delta V_3}{I_{3\_conv}} + \frac{C_{L\_conv} \Delta V_4}{I_{4\_conv}}.
\end{align}
\begin{align}
\label{eq:td_prop}
t_{d_prop} = \frac{C_{L\_prop} \Delta V_1}{I_{1\_prop}} + \frac{C_{L\_prop} \Delta V_2}{I_{2\_prop}} \nonumber\\
           + \frac{C_{L\_prop} \Delta V_3}{I_{3\_prop}} + \frac{C_{L\_prop} \Delta V_4}{I_{4\_prop}}.
\end{align}
where, $I_{i\_conv}$ and $I_{i\_prop}$ represent charging current in four different phases of conventional and proposed stage respectively, which are given by
\begin{gather}
I_{i\_conv}=I_{MP1}+I_{MP3}-I_{MN3}, \quad i=1,2,3,4.\\
I_{i\_prop}=I_{MP1}-I_{MN3}, \quad i=1,2,3,4.
\end{gather}
The equations and approximated average values of $I_{MP1}$, $I_{MP3}$ and $I_{MN3}$ in all phases are listed in Table \ref{tab:i_comparison}, where the charging process is divided into the four phases A, B, C and D as mentioned earlier. The regions of operation of the MOSFET are referred to as 0, 1 and 2 for cut-off, linear and saturation respectively.

The drain current of short-channel MOSFETs is assumed to follow the widely used alpha power law ~\cite{alpha_power_1}, ~\cite{alpha_power_2}. The carrier velocity saturation coefficient $\alpha$ is between $1.2$ to $1.5$ for sub-micron CMOS technology.
Then, according to the above equations and the estimated values in the table, we can get the propagation delay relationship between the proposed and conventional RO as $t_{d\_prop} = 88.4\%t_{d\_conv}$. Thus, the relation between the frequencies are:
\begin{align}
\label{eq:f_relation}
\frac{f_{prop}}{f_{conv}} = 1.156.
\end{align}
This means that the frequency of proposed RO increases $15.6\%$ compared with conventional structure for the same input current. From equation \ref{eq:f}, the average current relationship is expressed as:
\begin{align}
\label{eq:i_relation}
\frac{I_{prop}}{I_{conv}} = \frac{f_{prop}C_{L\_prop}}{f_{conv}C_{L\_conv}} = 0.867.
\end{align}
This implies the average current of the proposed RO decreases by $13.3\%$ compared to the conventional structure.

To verify these theoretical models, we conducted the current-frequency (I-F) transfer curve simulations of the conventional and proposed $4$ stage RO using $65$ nm CMOS models in SPICE. Fig. \ref{fig:i-f_sim} shows the simulation results and the theoretical prediction results of the output frequency. As predicted, the proposed oscillator indeed produces higher frequency than the conventional one for the same input current. However, the theoretical values are slightly greater than that of simulation because of the inaccuracy in estimation of the current of the transistor MP3.

\subsection{Robustness and Jitter}
While the increase in oscillation frequency is good, it is not useful if the new oscillator has a degradation in other key metrics such as robustness and jitter. We first evaluate the robustness of the proposed oscillator in comparison to the conventional one through simulations. The simulation results of line sensitivity ($\%/V$) and temperature sensitivity ($\%/T$) of four types of RO-CCO is shown in Fig. \ref{fig:line_senstivity}. As expected, the characteristics of differential current starved structures, both proposed and conventional, are better ($\approx 18\%$ for line sensitivity and $<0.014\%$ for temperature sensitivity) than non current starved pseudo differential RO ($\approx 139\%$ for line sensitivity and $\approx 0.14\%$ for temperature sensitivity). For current starved structures, differential structures are better than single ended ($\approx 61\%$ for line sensitivity and $\approx 0.06\%$ for temperature sensitivity) as expected. It can be seen that the line sensitivity and temperature sensitivity of the proposed differential structure is not degraded compared to the conventional differential structure. The relatively high value of sensitivity to power supply is traced back to the tail current sources coming out of saturation at power supply voltages lower than $0.9$V--this can be solved by reducing the current range. Confined to $1-1.2$V for power supply, the sensitivity is only around $1.5\%$.

Following \cite{Abidi_jitter} , the variance of period jitter for a RO can be expressed as:
\begin{align}
\label{eq:jitter_Abidi}
\sigma_{\tau} ^2 = \frac{KT}{If_{0}} (\frac{2}{V_{DD}-V_t} (\gamma_N + \gamma_P) + \frac{2}{V_{DD}} ).
\end{align}
where, 
\begin{align}
\label{eq:f0_Abidi}
f_0 \approx \frac{I/C}{NV_{DD}}.
\end{align}
$N$ is the number of delay stage. 
$\gamma_N$ and $\gamma_P$ are technology-dependent noise factors for NMOS and PMOS respectively.
Thus,
\begin{align}
\label{eq:jitter_propto}
\sigma_{\tau} ^2 \propto \frac{C}{I ^2}.
\end{align}
In our implementation, the load capacitor and average current decrease by $25\%$ and $13.3\%$ respectively.
\begin{align}
\label{eq:jitter}
\frac{\sigma_{Prop} ^2}{\sigma_{Conv} ^2} = \frac{(1-0.25)}{(1-0.133)^2} \approx 1.
\end{align}
So, the jitter of the proposed CCO is expected to be approximately the same as the conventional structure. This has been confirmed in simulations and measurement--we present these results in the next section.


\subsection{Frequency to Digital Conversion}
For usage within a neural network system, the raw frequencies of the CCO need to be often converted to a digital word. The easiest method to do this is to pass the CCO output as a clock to a counter\cite{elm_enyi}. However, this method incurs a large conversion time ($\propto 2^N$) and concomitantly large conversion energy. A different approach, following \cite{elmpuf_tcas1} is used in our work as described next.

\begin{figure}[!t]
\centerline
{\includegraphics[width=0.475\textwidth]{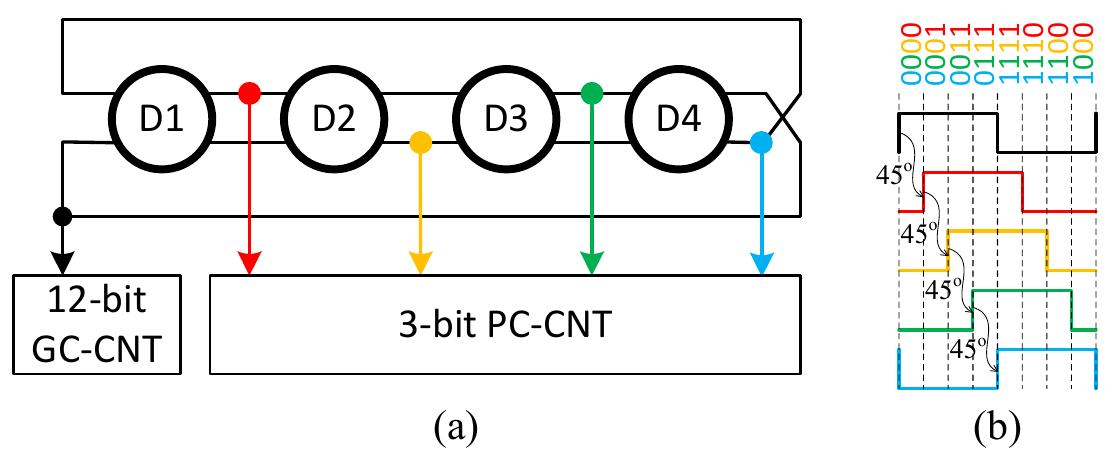}}
\caption{(a) Time to digital converter (TDC) composed of a four  delay stages differential CCO structure and two different counters for coarse and fine conversion separately. (b) The sequence diagrams for Gray code counter and the phase code counter with $45^\circ$ phase shift of each other as well as the corresponding 3-bit complementary phase code for one oscillation cycle.}
\label{counter_diagram}
\end{figure}

\begin{figure}[!htb]
\centerline
{\includegraphics[width=0.4\textwidth]{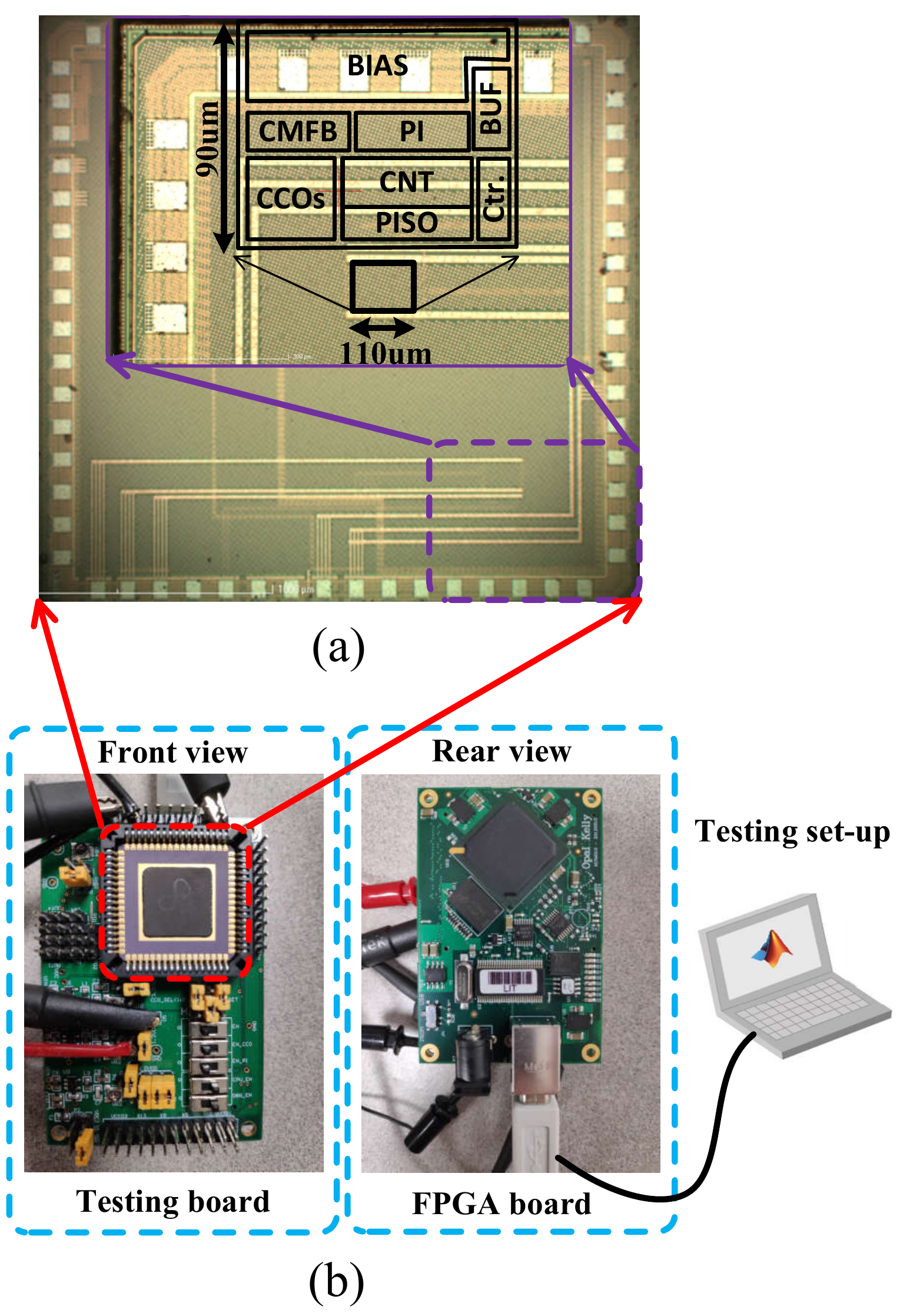}}
\caption{(a) Die photo and (b) testing set-up photo of the fabricated IC in $65$ nm CMOS. The testing board stacks with an FPGA board in charge of data transfer to and from the PC.}\label{fig:die}
\end{figure}

Fig. \ref{counter_diagram} shows the diagram of time to digital converter (TDC) with a four delay stage differential CCO and the sequential counters. Although the four stage differential CCO outputs eight phases totally, only half are unique with separate phase information while the rest are exactly the inverse of the unique phases and offer the same phase information. So only four phases with $45^\circ$ phase shift of each other, as shown in Fig. \ref{counter_diagram} (a), are utilized to generate the phase code counter (PC-CNT). Only $360^\circ /45^\circ =8$ codes are generated by the four stages differential CCO and hence only 3-bits are available in the PC-CNT acting as fine counter. The sequence diagrams of the four choice outputs and the corresponding complementary symmetry phase codes are illustrated in Fig. \ref{counter_diagram}(b). The opposite phase of the last phase of PC-CNT is chosen to clock a Gray code counter (GC-CNT) serving as coarse counter. The number of bits of GC-CNT depends on the input current range while the number of bits in PC-CNT determines the accuracy of the converter. A Gray code counter is selected for higher reliability in neural networks with tightly packed layout.
To enable wide dynamic range testing, a 12-bit GC-CNT is used in our design but the bit width can be optimized in neural network applications.

Both the phase code counter and Gray code counter possess the merits of energy efficiency and reliability since only one bit trips during state transition. Regarding the choice of number of stages in the CCO, we note that more stages of CCO structure ($\times M$) offers more valid output phases ($\times M$) and thus more bits in PC-CNT but has lower frequency and thus less bits in GC-CNT. Overall, the total number of bits is constant but more CCO stages consume  more area overhead ($\times M$). Assuming the same input current for both cases, the tail current for CCO core is constant and hence energy per conversion due to CCO is constant. Energy dissipated in the counter is less for more CCO stages, but this is much smaller than the CCO energy dissipation.
Hence, due to the need of small footprint of the CCO, a $4$-stage CCO core is designed in our work.

\section{Measurement Results}
\label{sec:results}



\begin{figure}[!t]
\centering
\subfigure[]{\includegraphics[width=0.45\textwidth]{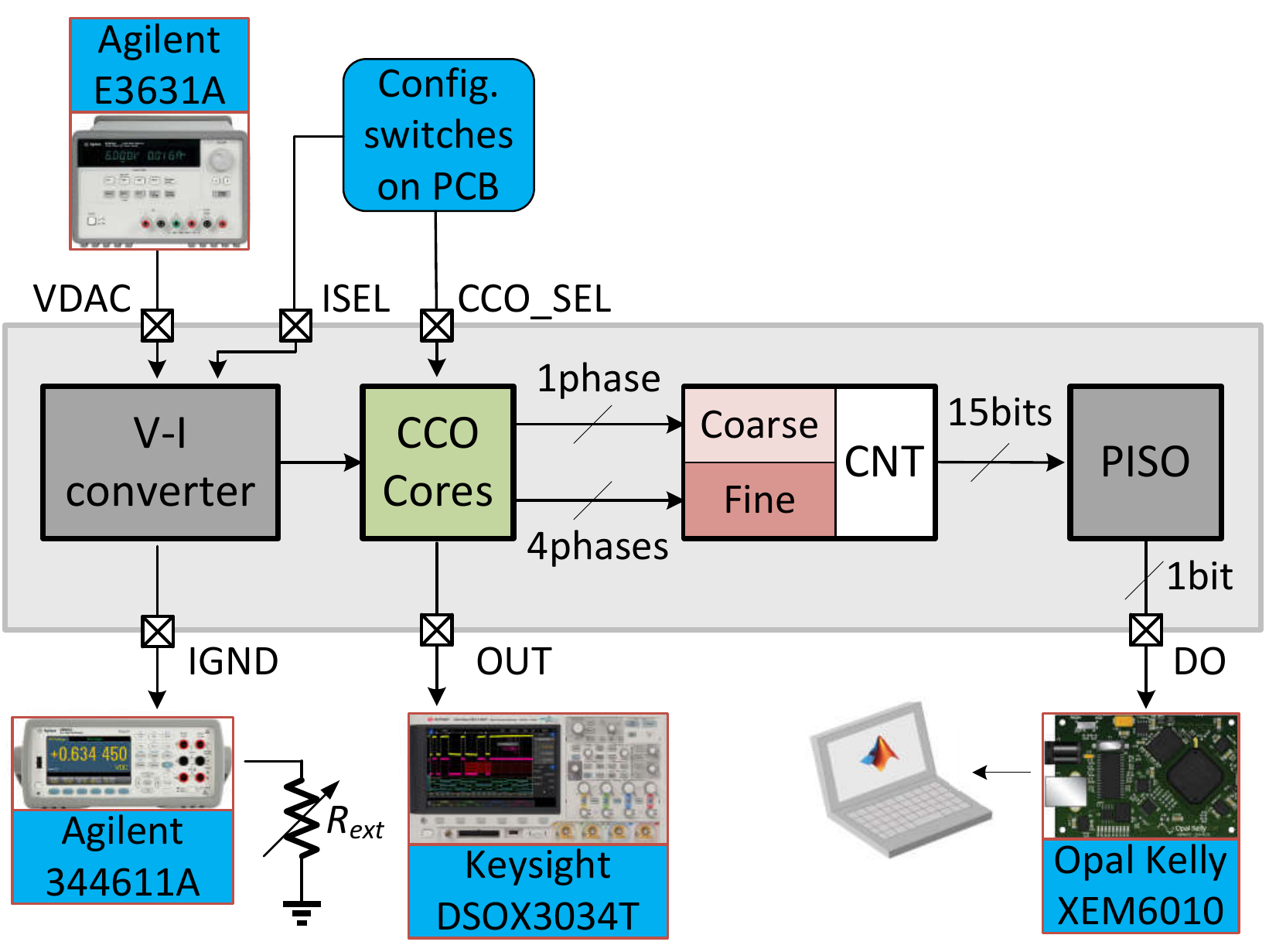}\label{block}}
\par\smallskip
\subfigure[]{\includegraphics[width=0.45\textwidth]{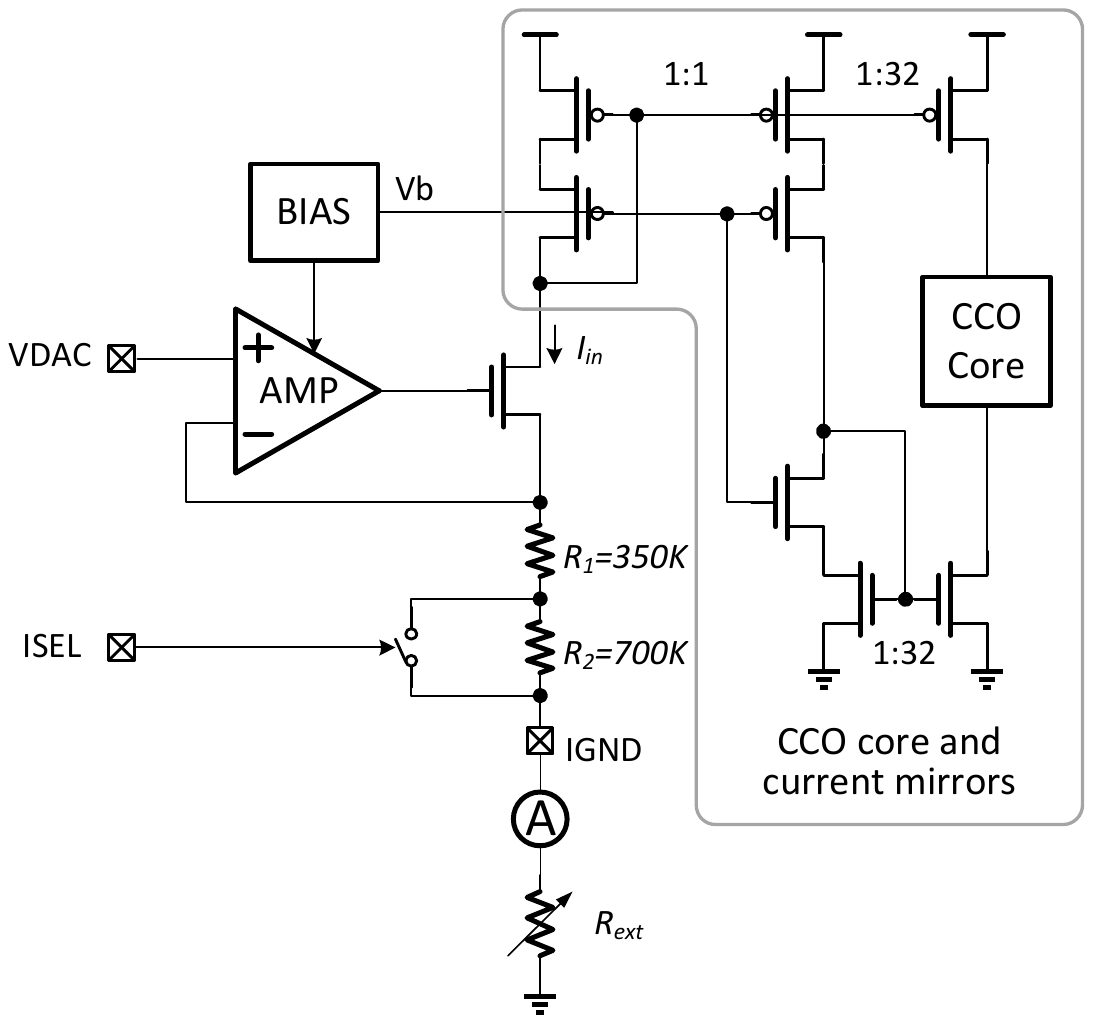}\label{V-I}}
\par\smallskip
\subfigure[]{\includegraphics[width=.498\textwidth]{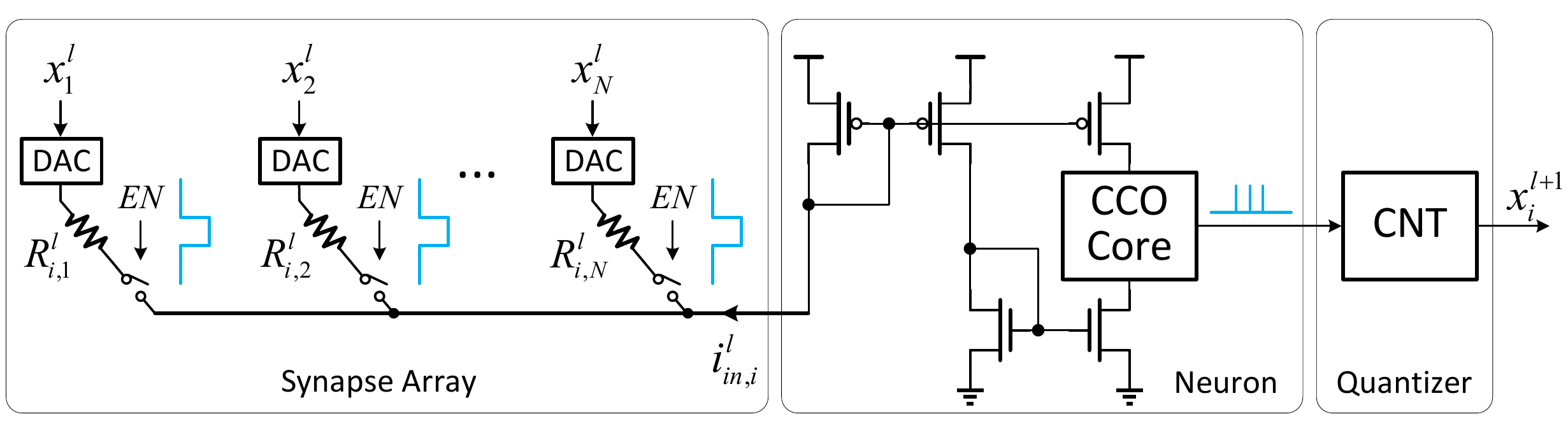}\label{circuit_snn1}}

\caption{(a) Block diagram of the fabricated IC and experimental setup for measurement. A tunable V-I converter provides input current to the CCO while a counter based on-chip digitizer is used to estimate the frequency of oscillation. (b) Circuit details of the V-I converter. ISEL is a digital bit used to enable or disable an on-chip resistor for coarse control of input current. The input current $I_{in}$ to the CCO will come from a synapse array in an actual neural network as shown in (c).}
\label{circuit_diagram}
\end{figure}

\begin{figure}[!htb]
    \centering
    \subfigure[]{\includegraphics[width=.4\textwidth]{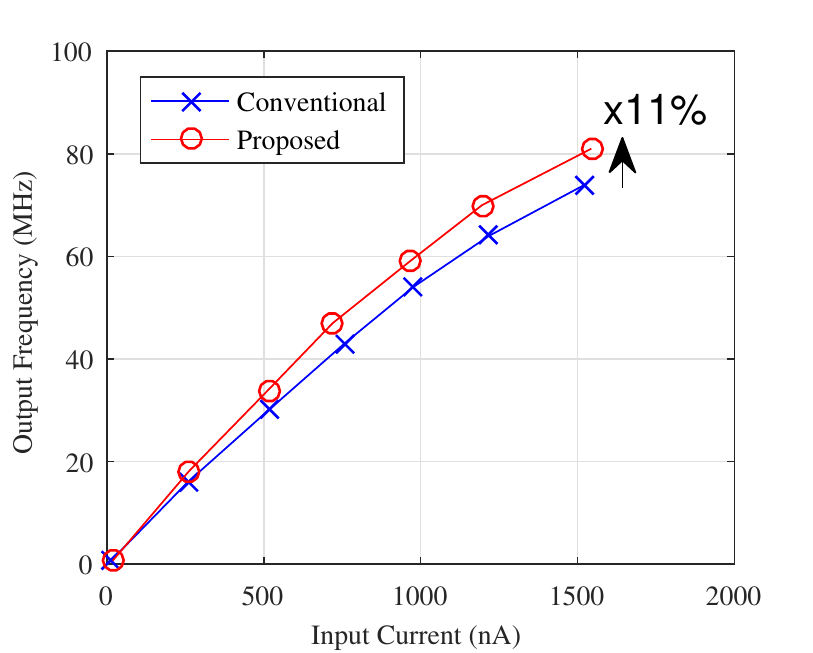}\label{Fig:f1}}\\
    \subfigure[]{\includegraphics[width=.4\textwidth]{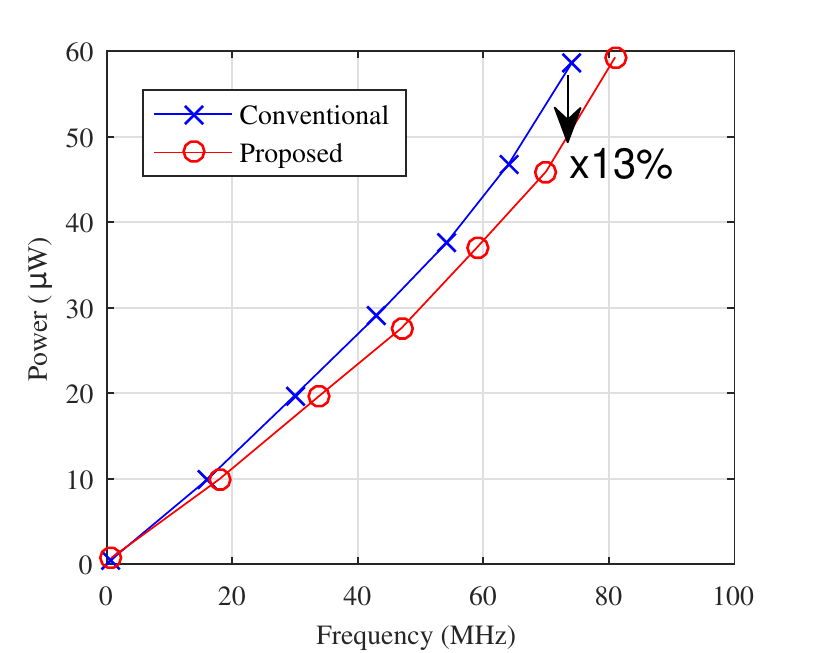}\label{Fig:f2}}\\
    \subfigure[]{\includegraphics[width=.4\textwidth]{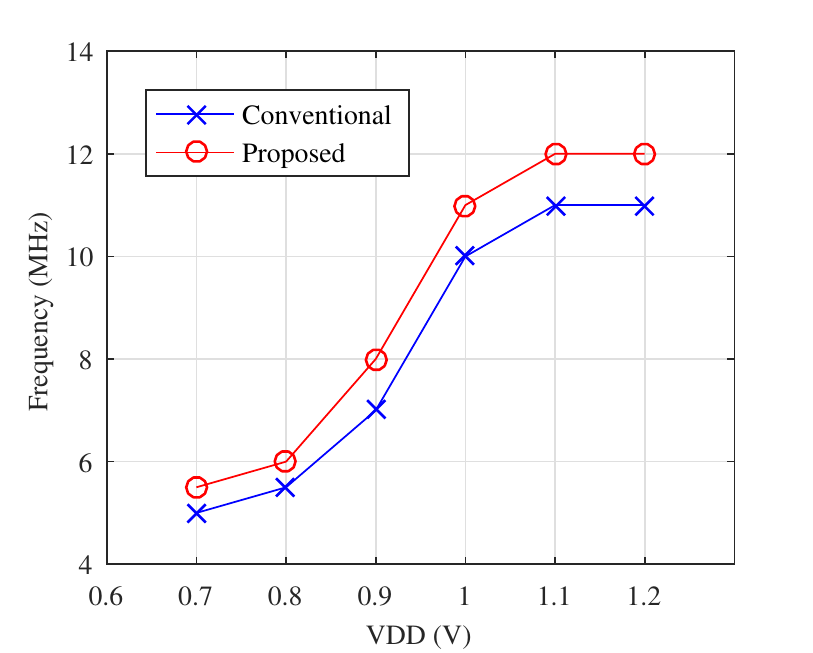}\label{Fig:f3}}
    \caption{
    \protect\subref{Fig:f1}~Measured frequency for different input current.
    \protect\subref{Fig:f2}~Measured power consumption at different frequencies.
    \protect\subref{Fig:f3}~Measured frequency for different $VDD$ under the condition of $VDAC=0.2$ V and $ISEL=0$.}
    \label{fig:i-f_measured}
\end{figure}

The prototype chip has been fabricated in a $65$ nm standard CMOS technology and includes a four-stage conventional CCO and the proposed CCO with output inverter buffers and peripheral circuits. The die micrograph and major sub-blocks are shown in Fig.\ref{fig:die}(a). The entire block is $110\times90um^2$ in size. The testing board is shown in Fig.\ref{fig:die}(b), where the configuration switches, testing points and  power supply are on the front side of the testing board while an FPGA board is hooked to the rear side. This section presents the measurement results and explains the testing condition and methodology. 

The testing setup is illustrated in Fig. \ref{circuit_diagram}(a). The bias current of the CCO can be controlled in coarse steps using an external resistor $R_{ext}$ on the board and in fine steps by a voltage input $VDAC$ from an on-board Digital to Analog Converter. The combination of resistor and voltage are converted into a reference current by a rail-to-rail operation amplifier based V-I converter (Fig. \ref{circuit_diagram}(b)) and then copied as the starved current for the CCO core by current mirrors. The range and amplitude of the current is also programmable by the configuration bit $ISEL$. CCO cores selection is configured by $CCO\_SEL$. The output of the oscillator is converted to a digital code by the coarse/fine digital quantizer as described earlier. The bit stream is serialized and sent to the FPGA for storage and analysis. The frequency of oscillation can be inferred from the digital code without having to measure a high frequency signal directly.

Note that the V-I converter is used just for convenience of testing. In an actual neural network, the input current $I_{in}$ to the CCO will come from a synaptic array. Fig. \ref{circuit_diagram}(c) shows an example how the CCO can be integrated into a neural network system with resistive synapses\cite{nature_joshua}. Here the CCO is used as the $i-th$ neuron in the $(l+1)-th$ layer and the count output from one neuron feeds a digital-analog converter (DAC) in the next layer to create voltage inputs. Note that this is just an example implementation just to clarify that the V-I converter is not a part of the neuron.

\begin{figure}[!t]
\centerline
{\includegraphics[width=0.45\textwidth]{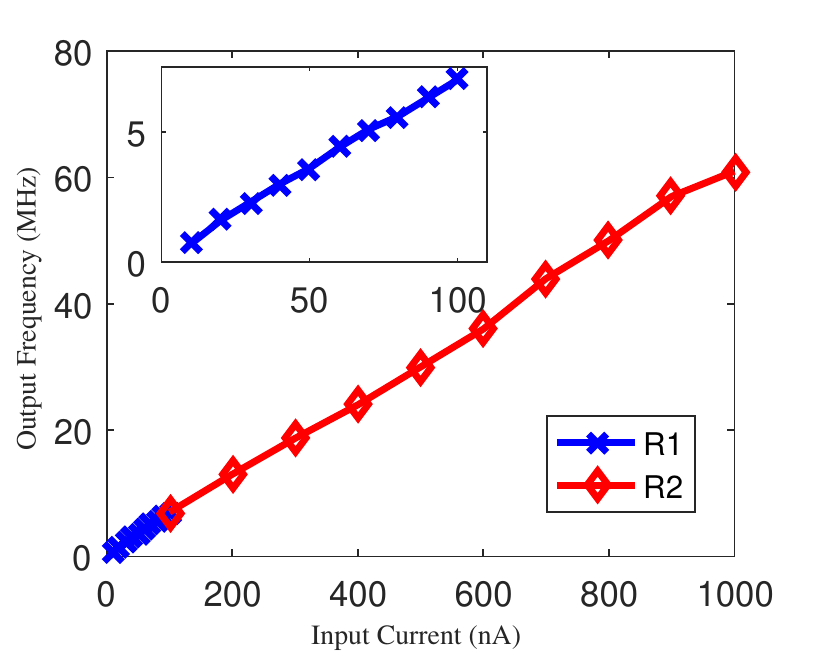}}
\caption{Measured frequency with different current by adjusting external resistor and input voltage respectively.}
\label{fig_res_vdac_f}
\end{figure}

\begin{figure}[!t]
    \centering
    \subfigure[]{\includegraphics[width=.45\textwidth]{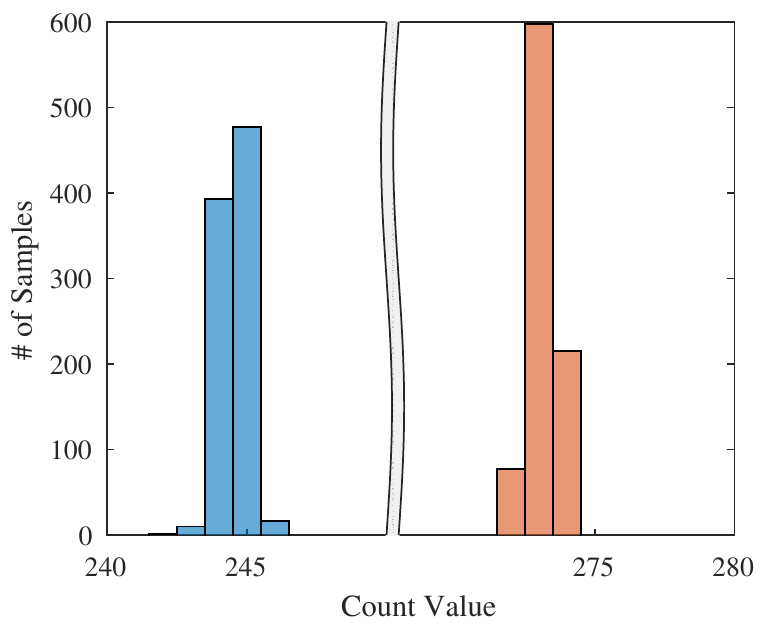}\label{Fig:j1}}\\
    \subfigure[]{\includegraphics[width=.45\textwidth]{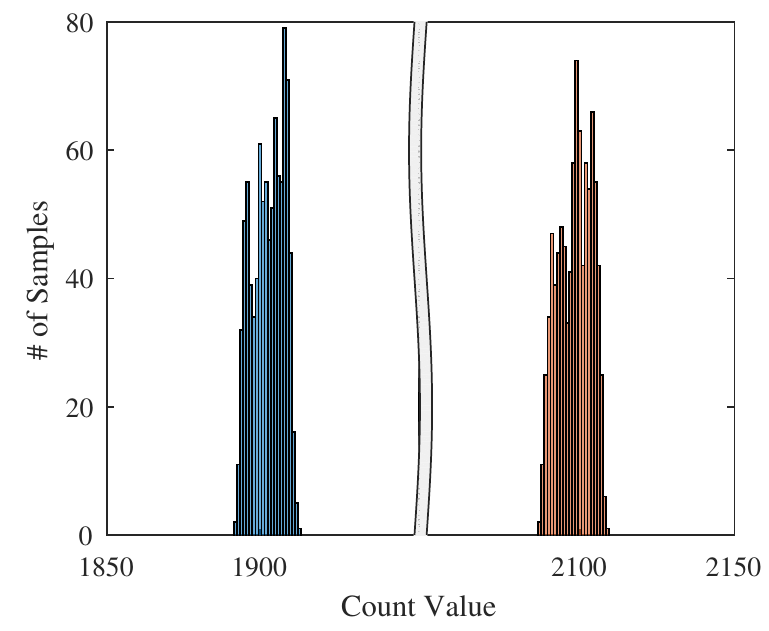}\label{Fig:j2}}
    \caption{Jitter performance comparison. The variation of the counter output for a specified time interval.
    \protect\subref{Fig:j1}~low input current, jitters of conventional structure (left) and proposed structure (right) are 0.23\% and 0.20\% respectively,
    \protect\subref{Fig:j2}~high input current, jitters of conventional structure (left) and proposed structure (right) are 0.26\% and 0.25\% respectively.}
    \label{Fig:jitter_measured}
\end{figure}


Measured frequency transfer curves of the conventional and proposed CCOs are plotted in Fig.\ref{fig:i-f_measured}(a) at different input current obtained by varying the DAC voltage over a wide range. The measured frequency of proposed CCO indeed achieves $\approx 11\%$ improvement compared with that of the conventional one for the same current input, ranging approximately from 0 to $1.5$ $uA$, close to the simulation based result in Section \ref{sec:prop}. 
Fig. \ref{fig:i-f_measured}(b) shows the measured power consumption at different oscillation frequencies. At the same oscillation frequency, the proposed CCO consumes $13\%$ less energy due to the less number of transistors and the corresponding capacitors. At the highest $V_{DD}=1.2$ V, the energy/conversion step including V-I converter ranges from $0.56-0.63$ $pJ/Cycle$. For a lower $V_{DD}=0.7$ V and excluding the V-I converter, the energy/conversion step ranges from $0.11-0.38$ $pJ/cycle$. Equivalently, this energy efficiency can be quoted as $0.11-0.38$ $\mu W/MHz$.
Finally, the dependence of the oscillation frequency on power supply voltage is tested. The experimentally obtained frequencies of the conventional and the proposed CCOs with different power supply $VDD$ are shown in Fig.\ref{fig:i-f_measured}(c). The testing is under the condition of $VDAC=0.2$ V, $R_{ext}=0$ $\Omega$ and $ISEL$ is set to low (total resistor is 1.05 $M\Omega$ in this case). The result shows a sharp drop in frequencies for $V_{DD}$ less than $1$ V and is traced back to the reference circuit not functioning properly at these voltages resulting in a change in the starving current. However, the proposed oscillator produce higher frequencies than the conventional design for all values of $V_{DD}$.




\begin{table}[!ht]
\centering
\caption{Characters and Performance Comparison of the Neuronal Oscillator}
\label{tab:performance_comparison_osc}
 \begin{threeparttable}
\begin{tabular}{@{}m{1.5cm}<{\centering}m{1.2cm}<{\centering}m{1.6cm}<{\centering}m{1.5cm}<{\centering}m{1.5cm}<{\centering}@{}}
\toprule[1.5pt]
Reference  & Technology ($nm$)  & Topology  & Power Eff. ($uW/MHz$) &Frequency ($MHz$)  \\ \hline
\midrule
This work & 65 & Differential, 4-stage  & 0.11-0.38\tnote{1}  & 80   \\\hline
Lee 2018 \\\cite{lee2018integrated} & 2000 (SOI) & Single-ended, 3-stage & 360	 & -  \\\hline
Yi 2019	\\\cite{elmpuf_tcas1} & 65 & Differential, 4-stage & 0.71  & 70  \\\hline
Yao 2017 \\\cite{elm_enyi} & 350 & Single-ended, relaxation & 0.2  & 4  \\\hline
Basu 2010 \\\cite{basu2010nullcline} & 350 & Single-ended, relaxation & 17.4	& 100  \\\hline
Sahoo 2017 \\\cite{TDNN_bibhu} & 65 & Single-ended & - & 1.5  \\
\bottomrule[1.5pt]
\end{tabular}
\begin{tablenotes}
\item[1] Excluding V-I converter power 
\end{tablenotes}
\end{threeparttable}
\end{table}

As mentioned before, the frequency of CCO can be tunable by both input voltage $VDAC$ and the external resistor $R_{ext}$. Fig.\ref{fig_res_vdac_f} shows the transfer curves, where $VDAC$ sweeps from 100 mV to 1 V. First, the external resistor $R_{ext}$ is fixed to $8.95$ $M\Omega$ and the configuration bit $ISEL$ is set to low, making the total resistance $R_{tot}=10$ $M\Omega$. Tuning input voltage $VDAC$ from 100 mV to 1 V covers the input current range from 10 nA to 100 nA, as shown in the blue line (R1).Then, the external resistor $R_{ext}$ is fixed to $650$ $K\Omega$ and $ISEL$ is set to high, leading the total resistance $R_{tot}=1$ $M\Omega$.Tuning input voltage $VDAC$ from 100 mV to 1 V covers the input current from 100 nA to 1000 nA, as shown in the red line (R2).  The testing points show the two kinds of tuning matches well and one can act as coarse tuning while another as fine tuning.

\begin{table*}[ht!]
\centering
\caption{Performance Summary and Comparison with State of the Art}
\label{tab:performance_comparison}
\begin{threeparttable}

\begin{tabular}{@{}m{1.5cm}<{\centering}m{1.2cm}<{\centering}m{1.6cm}<{\centering}m{1cm}<{\centering}m{1.2cm}<{\centering}m{0.8cm}<{\centering}m{1.2cm}<{\centering}m{1cm}<{\centering}m{3.6cm}<{\centering}@{}}


\toprule[1.5pt]
Reference  & Technology ($nm$)  & Architecture  &Supply ($V$)  & Power ($mW$)   &  Area ($mm^2$) &Bandwidth ($MHz$) & FOM ($fJ/conv.$) & Application Comments \\ \hline
\midrule
This work & 65 & Novel CCO + TDC &1.2 & 0.02   & 0.004  & 0.5 & 79\tnote{1} &  silicon neuron in neuromorphic applications \\\hline
Leene 2018 \\\cite{leene20180} & 65 & VCO + $\Delta\Sigma$ & 0.5 & 1.3  & 0.006  & 0.011 & 175 & electrode bio-potential recordings  \\\hline
Li 2017	\\\cite{li2017174} & 130 & VCO + $\Delta\Sigma$ & 1.2 & 1.05   & 0.13 & 0.4 & 118 & continuous-time delta-sigma modulator \\\hline
Young 2014 \\\cite{young201475db} & 65 & OTA + VCO & 1.2 & 38   & 0.49 & 50 & 294 & continuous-time delta-sigma modulator \\\hline
Talor 2013 \\\cite{taylor2013reconfigurable} & 65 & VCO & 0.9	& 11.5   & 0.075 & 5.08 &246 & digital wireless receiver, CT delta-sigma modulator  \\\hline
Kim 2014 \\\cite{kim201411} & 130 & delay cell + 3-D Vernier & 1.5 & 0.33     & 0.28 & 0.5 & 400 & time-of-flight (ToF) application\\\hline
Tu 2017 \\\cite{tu2017low} & 40 & PWM + $\Delta\Sigma$ & 1.2 & 0.02   &  0.015 & 0.005 &1643 & CMOS Image Sensor, X-ray detector \\\hline
Jayaraj 2019 \\\cite{jayaraj2019highly} & 65 & VCO + $\Delta\Sigma$ & 1.2 & 1    & 0.06 & 2.5 & 151 & continuous-time (CT) ADC \\ \hline
Jayaraj 2019	\\\cite{Jayaraj_ASSCC19} & 65 & VCO + $\Delta\Sigma$ & 1 & 0.1 & 0.06 & 2.3 & 8.6 & continuous-time (CT) ADC  \\\hline
Zhong 2018 \\\cite{Zhong_ASSCC18} & 40 & VCO + $\Delta\Sigma$ & 1.1 & 0.91 & 0.086 & 5.2 &- & continuous-time (CT) ADC  \\
\bottomrule[1.5pt]
\end{tabular}
\begin{tablenotes}
\item[1] FOM is based on simulation since V-I converter does not support high frequency input.
\end{tablenotes}
\end{threeparttable}
\end{table*} 


To test the jitter performance, the outputs of CCOs are connected to a time-to-digital converter (TDC) in the chip, as shown in Fig. \ref{circuit_diagram} and described in the earlier section. A parallel in serial out (PISO) circuit  is used to reduce the number of output pads. The 15-bit parallel data is loaded into the register simultaneously and is shifted out of the register serially one bit at a time under clock control. In our testing, the serial output data is samples by FPGA and then passed to PC to decode. Fig. \ref{Fig:jitter_measured} (a) and (b) display the measured {period} jitter performance for the conventional and proposed structure at a low and high frequency respectively. {The duration of counter window is set to 50 $\mu s$ in the measurement.} It can be seen that the two structures perform approximately the same with around 0.2\% jitter at lower frequency and around 0.25\% jitter at higher frequency.

\begin{figure}[!t]
    \centering
    \subfigure[Area]{\includegraphics[width=.24\textwidth]{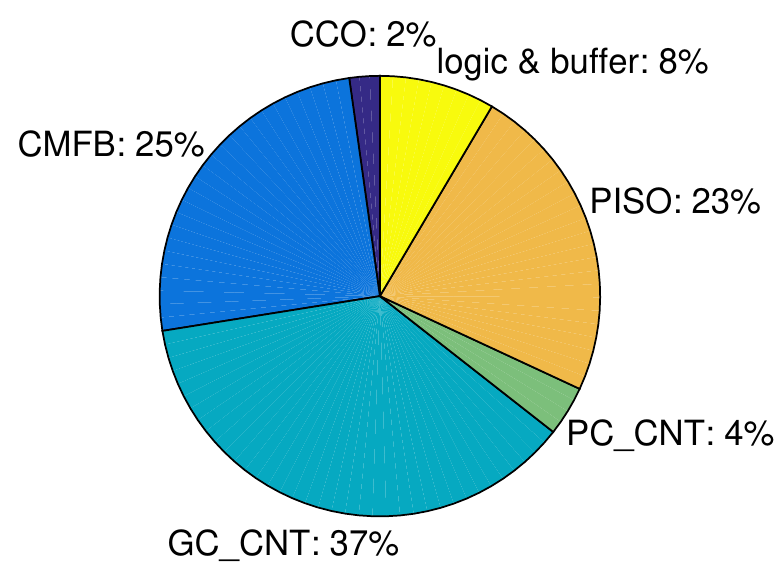}\label{pie_area}}
    \subfigure[Power]{\includegraphics[width=.24\textwidth]{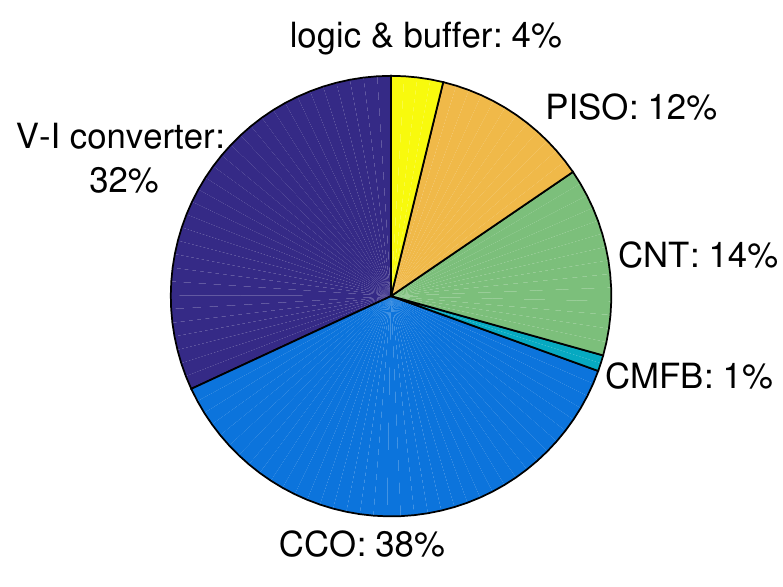}\label{pie_power}}
    \caption{(a) Area and (b) Power contributions (simulation) from each sub-circuit in the system.
    }
    \label{fig_pie}
\end{figure}

\begin{figure}[!t]
\centerline
{\includegraphics[width=0.475\textwidth]{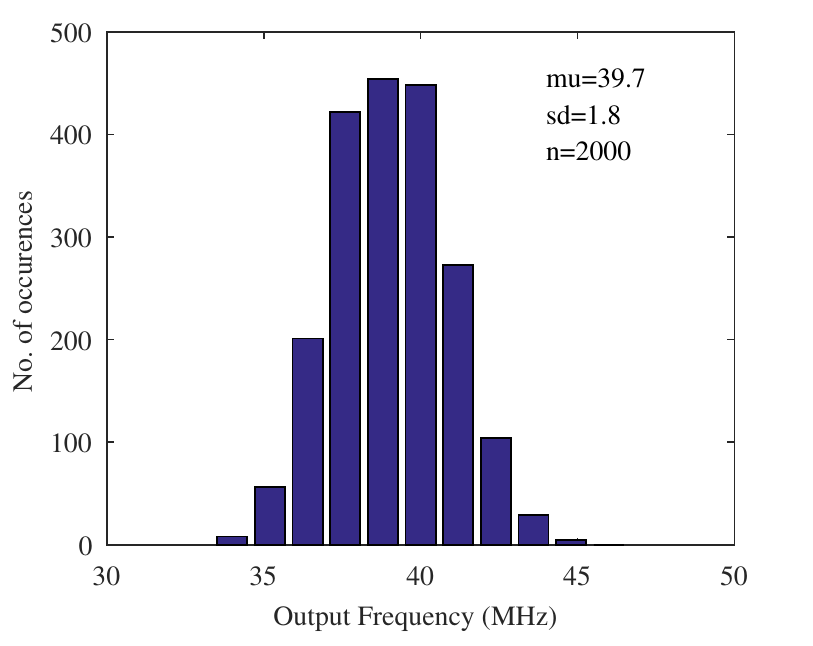}}
\caption{Monte Carlo simulation of the proposed CCO generated from $2000$ runs.}
\label{fig:MC_CCO}
\end{figure}

Fig. \ref{fig_pie} shows the relative area utilization and the simulated power breakdown. The 4-stage CCO core contributes a small area of $44$ $um^2$. It may be noted that although the V-I converter consumes a significant system power in our testing, this block is not an intrinsic part of our design but is there only for testing as mentioned earlier.


Table \ref{tab:performance_comparison_osc} compares characteristics and performance of various oscillators used in neuromorphic applications. The proposed CCO has one of the highest energy efficiencies reported so far (except \cite{elm_enyi}). However, \cite{elm_enyi} presents a single ended design with a much lower frequency of operation. Further, compared with the other differential design in \cite{elmpuf_tcas1}, the proposed work uses $25\%$ less area.

Further, a detailed performance comparison is summarized in Table \ref{tab:performance_comparison} for the recent time based ADC architectures. To compare with other ADCs, the center frequency of CCO is set to be $60$ MHz in the prototype ADC. With the input analog bandwidth of $500$ kHz, running at $5$ MS/s, it consumes $72$ $\mu W$ from $1.2$ V in average with full input signal swing, out of which $40$ $\mu W$ is drawn by the digital circuits and the buffer while $32$ $\mu W$ is from analog (including CCO core and bias).
The proposed novel CCO structure consumes less area and lower energy overhead by using less transistors. Besides, the Gray code counter and phase code counter consume less dynamic energy due to their merit of minimum change code. These lead to high energy efficiency of the time based ADC quantified by the figure of merit (FOM) calculated as:
\begin{align}
\label{eq:fom}
FOM = \frac{P}{2^{ENOB}*2*BW}
\end{align}
Where, $P$ is the total power consumed by the CCO-based ADC.
The ADC achieves an FOM in the range of $79–102$ fJ/conv-step for different input frequencies.
{Note that \cite{Jayaraj_ASSCC19} achieves much lower FOM by using $2^{nd}$ order noise shaping loop; however, this benefit comes mostly due to reduced quantization noise and is orthogonal to the improvements in oscillator structure we report. Also, this improvement comes with a $>10$X area penalty which may not be suited for neuron designs.}

\section{Discussion}
\label{sec:discussion}
\subsection{Neural Network Simulation}
\label{sec:NNSIM}


In this section, we explore potential application of the proposed CCO as a neuron in neural network (NN) implementations. The custom activation function of the proposed neuron is modelled using experimental results of the CCO (Fig. \ref{fig:i-f_measured} and Fig. \ref{Fig:jitter_measured}). The I-F transfer curve of CCO has a high sensitivity to process variations and hence the neurons on a chip are expected to have mismatch. Fig. \ref{fig:MC_CCO} shows the mismatch of CCO through Monte Carlo (MC) simulation. As seen from the figure, the standard variation is $1.8$ MHz at the mean frequency of $39.7$ MHz leading to a coefficient of variation of $\approx 5\%$. The mismatch of CCO is observed by performing MC runs while sweeping input currents. The variability in the slope of this I-F curve is included in the NN simulation as shown next. \textcolor{black}{Note that synaptic non-idealities are a separate issue and has to be considered for the full system as well. This is beyond the scope of this work but many strategies to handle synaptic non-idealities such as mismatch, low-precision and write non-linearity have been published\cite{elm_enyi,opto-natcomm,bhaduri_neco,sroy_tnnls,querlioz}.}

At first, the neural networks are trained and tested using ReLu activation function to establish a benchmark for further evaluation of proposed activation function. For hardware simulation, we adopted two strategies. In the first one, the networks are trained using ReLu activation and the custom activation function is introduced only during inference. In the second strategy, the custom activation is used both during training and inference. The simulation details are described in detail in algorithm \ref{Algo: neuron_model}.

\begin{algorithm}
\caption{Hardware Simulation}
\label{Algo: neuron_model}
\begin{algorithmic}[1]
\STATE Fit a straight line for each set of readings  (i-f curves) and get a distribution of slopes. Calculate mean slope $\tilde{m}$.
\STATE Divide the curves in two regions and fit a straight line for each region. Optimum boundary between regions $\mathcal{B}$ is determined based on minimising MSE over all curves.
\STATE For both the regions, obtain the slope distributions from Monte-Carlo simulations and model them using Gaussian distributions: $\mathcal{N}_1(\mu_1,\sigma_1^2)$ and $\mathcal{N}_2(\mu_2,\sigma_2^2)$
\STATE Noise is modelled as Gaussian noise $n \sim \mathcal{N} (\mu=0,\sigma=jitter)$
\STATE Finally the custom activation function is given by: Custom Activation(x) = $ReLu(m_1 x [x<\mathcal{B}]+m_2 x[x>\mathcal{B}])  +  (m_1 x [x<\mathcal{B}] +m_2 x[x>\mathcal{B}])\times n$, where $m_1 \sim \mathcal{N}_1$ and $m_2 \sim \mathcal{N}_2$
\STATE \textbf{For hardware simulation (inference):}\par
\hskip\algorithmicindent training activation: $ReLu(\tilde{m}x)$ \par
\hskip\algorithmicindent inference activation: Custom Activation(x)
\STATE \textbf{For hardware simulation (training + inference):}\par
\hskip\algorithmicindent training activation: Custom Activation(x)\par
\hskip\algorithmicindent inference activation: Custom Activation(x)
\end{algorithmic}
\end{algorithm}

\begin{table*}[!htbp]
\centering
\caption{Neural Network Simulation}
\label{tab: NNsim}
\begin{tabular}{|c|c|c|c|c|c|c|c|}
\hline
\multicolumn{2}{|c|}{Datasets}                                                                                                                            & \multicolumn{2}{c|}{MNIST} & \multicolumn{2}{c|}{Fashion MNIST} & \multicolumn{2}{c|}{CIFAR-10} \\ \hline
\multicolumn{2}{|c|}{Network}                                                                                                                             & ANN          & CNN         & ANN              & CNN             & ANN           & CNN          \\ \hline
\multirow{3}{*}{\begin{tabular}[c]{@{}c@{}} \\Simulation\\ Accuracy\end{tabular}} & Software                                                                 & 0.982        & 0.9926      & 0.8919           & 0.9174          & 0.5284        & 0.7478       \\ \cline{2-8} 
                                                                              & \begin{tabular}[c]{@{}c@{}}Hardware \\ (inference)\end{tabular}          & 0.9812       & 0.9915      & 0.8846           & 0.9088          & 0.5014        & 0.6974       \\ \cline{2-8} 
                                                                              & \begin{tabular}[c]{@{}c@{}}Hardware\\  (training+inference)\end{tabular} & 0.9829       & 0.9931      & 0.8937           & 0.9135          & 0.5257        & 0.7435       \\ \hline
\end{tabular}
\end{table*}

\begin{figure*}[!t]
    \centering
    \subfigure[] {\includegraphics[width=.3\textwidth]{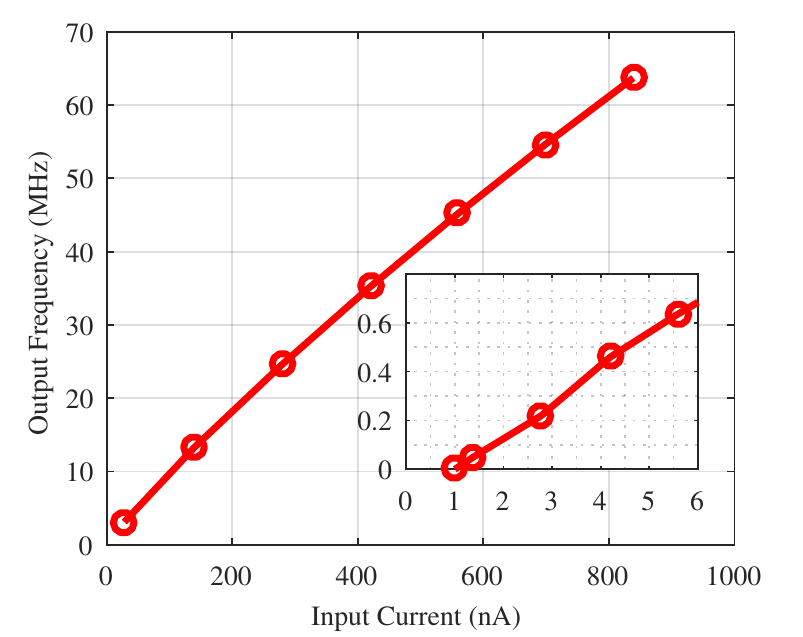}\label{snn1}}
    \subfigure[] {\includegraphics[width=.33\textwidth]{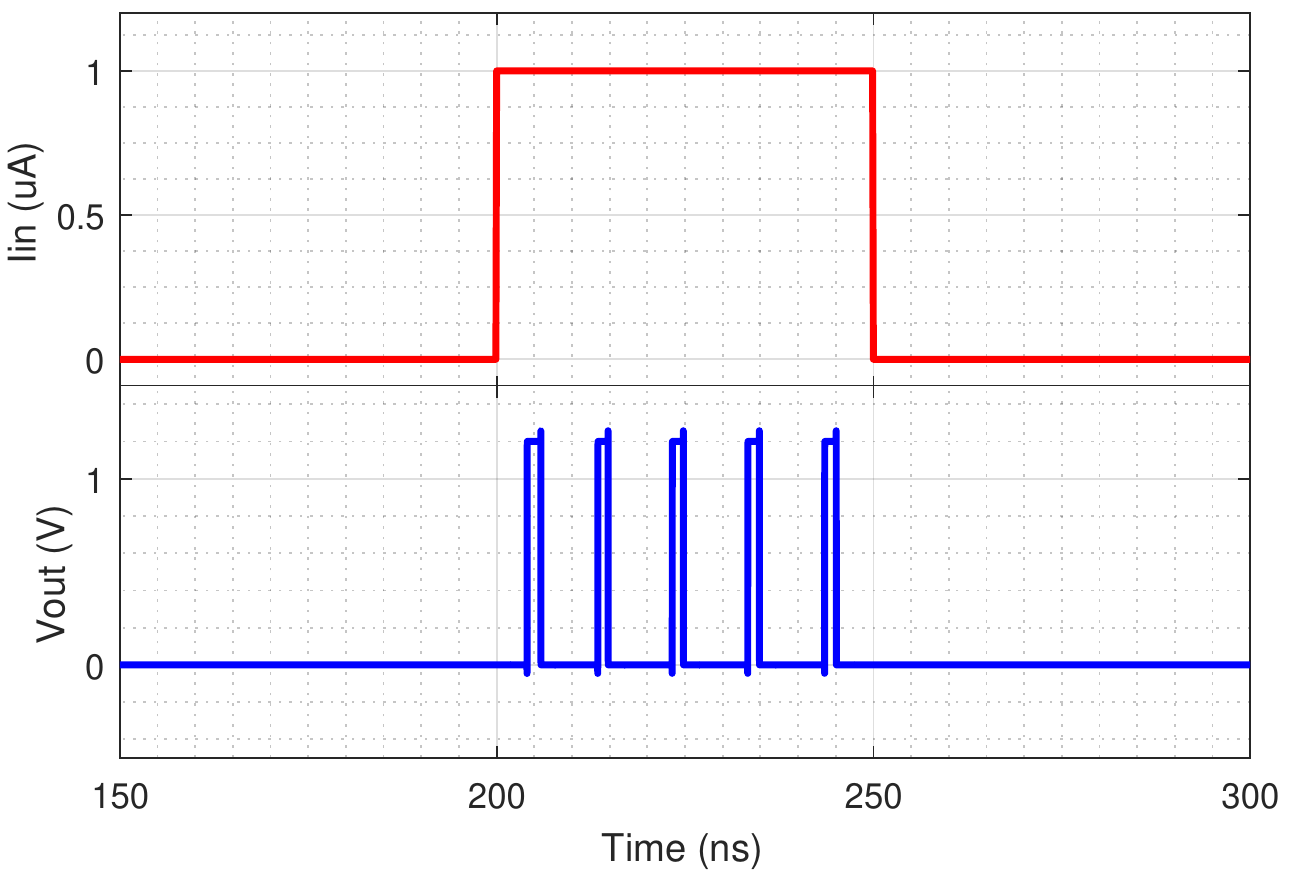}\label{snn2}}
    \subfigure[] {\includegraphics[width=.33\textwidth]{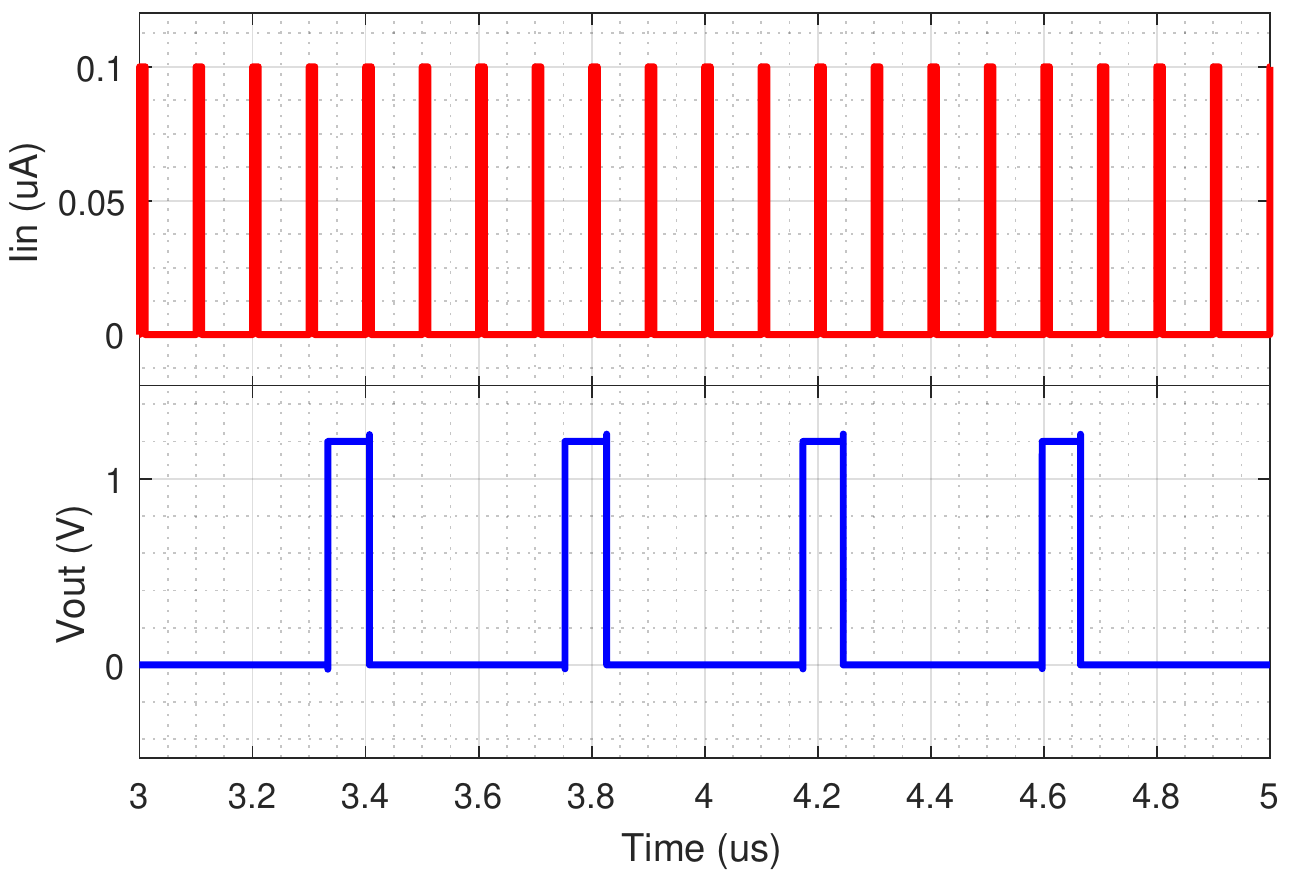}\label{snn3}}
    \caption{Usage of the CCO in an SNN based on SPICE simulations: (a) With leak current of $1$ nA added, the CCO does not oscillate for input current less than $I_{leak}$. (b) For a step current input, the neuron only oscillates during the step input. (c) For pulse current inputs, the charge is integrated on a current mirror's capacitor and outputs spike when the CCO reaches a desired phase.}
    \label{fig:snn}
\end{figure*}

We used three different datasets (MNIST~\cite{mnist}, fashion MNIST~\cite{FMNIST}, CIFAR-10~\cite{CIFAR10}) and two network topologies (one ANN and one CNN) to evaluate the performance of the proposed neuron model. Since the primary aim of these simulations is to verify the viability of the CCO based neuron, instead of experimenting with different network architectures, we used the same ANN and CNN architectures for all the datasets and compared the performance of the hardware simulations with their pure software counterpart. The ANN comprises of two fully connected hidden layers with 800 and 300 neurons respectively. The CNN network architecture used can be described as: $A\times B\times C-32c5-m2-64c5-m2-128c1-m2-fc512-0.5d-fc10$, where the input dimension is $A\times B\times C$, $XcY$ represents a convolution layer with $X$ convolution filters with $Y\times Y$ size, $mZ$ represents a $Z\times Z$ 2D max-pooling layer, $Pd$ represent a dropout layer with dropout rate $P$ and $fcL$ represent a fully connected dense layer with $L$ neurons. All models are trained and tested through google colaboratory on a server with Intel Xeon CPU and NVIDIA Tesla K80 GPU. All the models are trained with categorical crossentropy loss and Adam optimizer and the testing accuracies for all the datasets and network topologies are averaged over 3 trials.

 The results (\textcolor{black}{mean across 3 trials}) are shown in table \ref{tab: NNsim}. The key observations from the results are as follows: firstly, for simple datasets like MNIST, there is almost no loss of accuracy even if the custom activation function is introduced only during inference. For a relatively more complex fashion MNIST dataset, there is a small loss ($\sim 1\%$) in testing accuracy, while for more complex CIFAR10 dataset, the loss of accuracy becomes significant ($\sim 2.5-4\%$). Secondly, the loss in accuracy can be almost completely recovered even for complex datasets like CIFAR-10, if the custom activation is introduced during training i.e. weight learning through back-propagation is able to correct for any non-ideality resulting from the custom activation. Thirdly, in some cases, when the custom activation is introduced during training, the testing accuracies seem to surpass the software accuracy. This can be attributed to Gaussian Noise injection in the custom activation acting as noise regularization and improving overall network performance~\cite{noiseregular}. For all the datasets and both hardware simulation settings, the loss of accuracy is similar for both ANN and CNN network topologies. Since the ANN uses only two hidden layers while the CNN architecture consists of large number of layers, this observation establishes the scalability of the proposed neuron model. \textcolor{black}{Finally, we evaluated in detail the variability in the performance of the classifiers caused by random variation of the custom activation function described in step $5$ of algorithm \ref{Algo: neuron_model}. For MNIST classification task with ANN and CNN, we evaluated the classification accuracies over $20$ trials where the slopes of the custom activation are randomly generated for each trial. The standard deviation of classification accuracy for ANN and CNN are $0.15\%$ and $0.10\%$ respectively. This goes to show that the performance variation introduced by the slope variability is very small and therefore, further proves the viability of the proposed neuron model.}
 
\begin{figure*}[!t]
\centerline
{\includegraphics[width=0.9\textwidth]{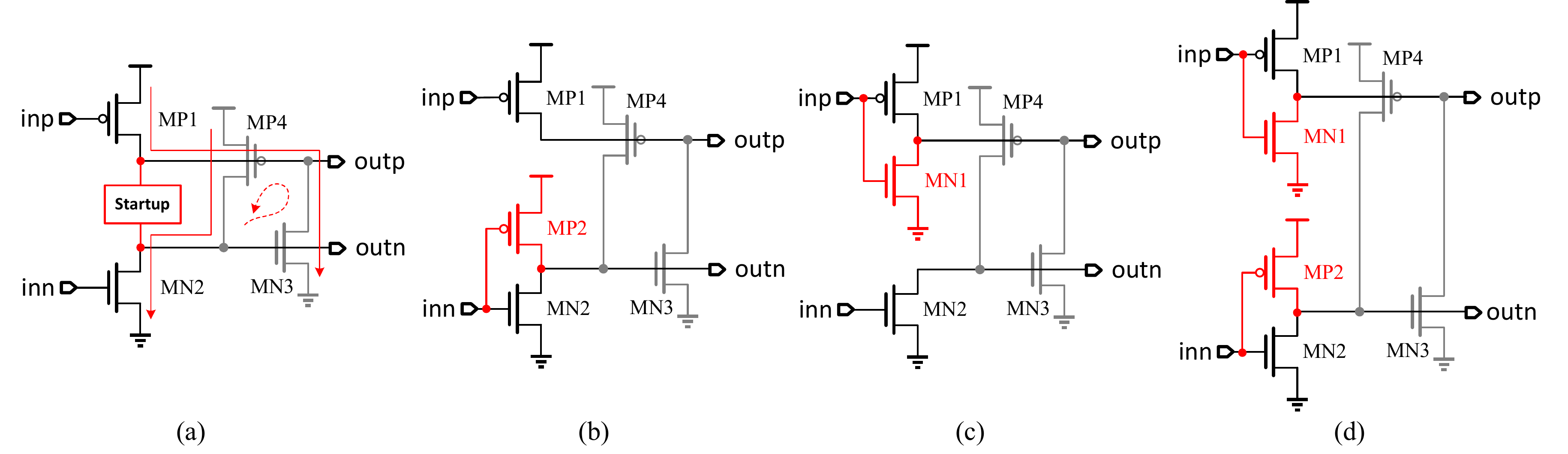}}
\caption{ Diagram of a single delay stage with (a) start-up circuit. (b) a PMOS start-up. (c) a NMOS start-up. (d) both PMOS and NMOS start-up.}
\label{fig:startup_circuits}
\end{figure*}

Changes in power supply voltage and temperature are ideally common mode perturbations and affect the CCO less due to differential structure. Moreover, due to sharing of power supply, if all neurons are affected by the same factor, RELU based neural networks will not be ideally affected. However, due to variations between neurons, there will be some effect in real implementations. To capture this effect, Monte-Carlo simulations were done to find distributions of supply voltage sensitivity (temperature sensitivity is almost negligible as shown in Section \ref{sec:prop}). The resulting sensitivity distribution had a mean of $-0.78\%$ and standard deviation of $1.74\%$ when $V_{dd}$ varied in the range of $1-1.2$ V. Including this as an additional perturbation for the neuron outputs during inference, the results for CIFAR-10 were re-evaluated. We observed negligible reduction in accuracy of $<0.5\%$ confirming our hypothesis that the network is relatively robust against supply variations.

\subsection{CCO as Spiking Neuron}
\label{sec:CCO_SNN}
The earlier sections showed the usage of the CCO as a rate encoded neuron similar to its use in \cite{elm_enyi,elmpuf_tcas1}. However, it may also be used as a spiking neuron with precisely timed output spikes as needed in spiking neural networks (SNN). While one way to use the CCO as a spiking neuron is shown in \cite{TDNN_bibhu}, we show next how some of the critical features can be included in the CCO much easily. We identify one of the phases of the CCO as the spike phase and the corresponding output of the fine counter is plotted as spike output.

First, we show in SPICE simulations that adding a leak current source\cite{giacomo_neuron} next to the PMOS input current mirror in Fig. \ref{V-I} can act to suppress spurious neuron spikes at low input currents. The resulting I-F curve plotted in Fig. \ref{fig:snn}(a) shows that indeed valid neuron firing starts for input currents larger than $I_{leak}=1 nA$. Next, the response of the CCO neuron to a step of magnitude $I_{in}=1 uA$ is shown in Fig. \ref{fig:snn}(b). As expected, the neuron fires spikes only when the step is applied. Lastly, to see how to integrate such a neuron in a network, we show the response of the CCO neuron to spike inputs that may come from other neurons in the network. Plotted in Fig. \ref{fig:snn}(c), it shows that for input synaptic current pulses of magnitude $100 nA$, the neuron integrates the charge on the PMOS current mirror and produces an output spike for every $4$ input spikes.
\subsection{Startup Circuit}
\label{sec:CCO_Startup}

\begin{figure}[!t]
\centerline
{\includegraphics[width=0.45\textwidth]{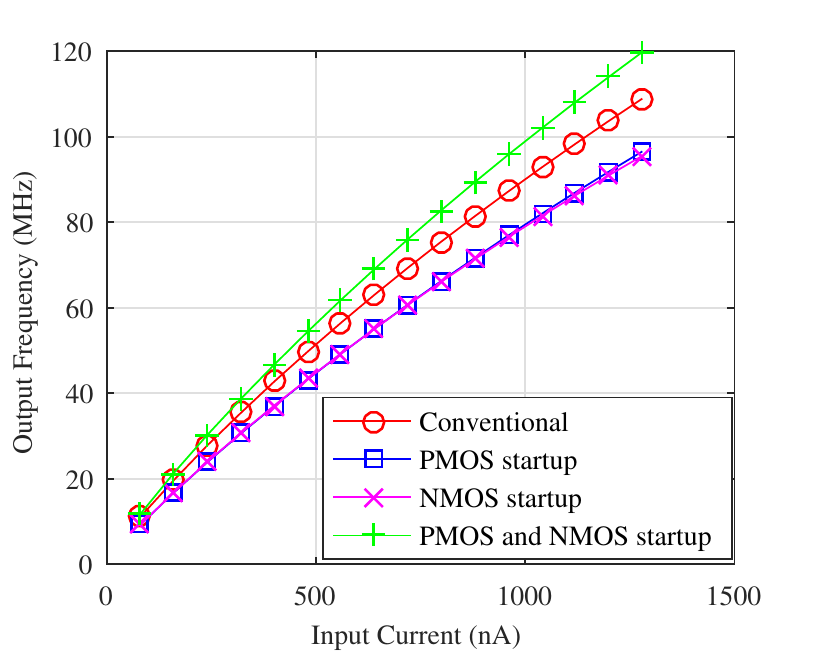}}
\caption{Comparison between the frequency of CCOs with different start-up circuits.}
\label{fig:f_startup}
\end{figure}

In Section \ref{sec:prop}, we proposed the novel structure of CCO and explained the necessity of a start-up circuit. A robust oscillation can be ensured by a start-up circuit with only a PMOS transistor or a NMOS transistor or both. As shown in Fig. \ref{fig:startup_circuits} (a), a start-up circuit is equipped at the differential output nodes to eliminate the uncertainty state. The start-up circuit would be a single PMOS acting as a pull-up device to eliminate the uncertainty of $outn$, as shown in Fig. \ref{fig:startup_circuits} (b), or a single NMOS acting as a pull-down device to eliminate the uncertainty of $outp$, as shown in Fig. \ref{fig:startup_circuits} (c), or both a PMOS and a NMOS to eliminate the uncertainty of both the differential outputs, as shown in Fig. \ref{fig:startup_circuits} (d). 

The different start-up circuits give different oscillation frequencies. Fig. \ref{fig:f_startup} compares the simulated output frequencies of conventional and proposed structure with three different start-up circuits. The CCO with both PMOS and NMOS start-up performs $\times 11\%$ higher frequency than the conventional one while the CCOs with a single PMOS or NMOS start-up degrade the frequencies around $\times 13\%$ compared to the conventional one. Hence, we only use the startup with both PMOS and NMOS in the proposed structure.

\section{Conclusion}
\label{sec:conclusion}
In this paper, an energy- and area-efficient full differential CMOS current controlled ring oscillator (CCO) is presented as a suitable and compact structure in neurocomputing applications. The CCO  achieves higher frequency while consuming lower area and lower energy due to less transistors being utilized compared with the conventional structure. By eliminating the unnecessary transistors, the proposed structure is composed of a simplest dynamic positive feedback latch and differential pairs, saving $\approx 25\%$ in size.
The CCO can be tuned by both input voltage and external variable resistor. The measurement results show our work achieves 11\% frequency improvement and 13\% energy-efficient without degrading the jitter and phase noise characteristics. Simulations of both ANN and CNN with the proposed CCO characteristics as the neuron transfer curve show no degradation in accuracy if the CCO nonlinearity is included in training.

\Urlmuskip=0mu plus 1mu\relax
\bibliographystyle{IEEEtran}
\bibliography{References.bib}

\begin{thebibliography}{10}
\providecommand{\url}[1]{#1}
\csname url@samestyle\endcsname
\providecommand{\newblock}{\relax}
\providecommand{\bibinfo}[2]{#2}
\providecommand{\BIBentrySTDinterwordspacing}{\spaceskip=0pt\relax}
\providecommand{\BIBentryALTinterwordstretchfactor}{4}
\providecommand{\BIBentryALTinterwordspacing}{\spaceskip=\fontdimen2\font plus
\BIBentryALTinterwordstretchfactor\fontdimen3\font minus
  \fontdimen4\font\relax}
\providecommand{\BIBforeignlanguage}[2]{{%
\expandafter\ifx\csname l@#1\endcsname\relax
\typeout{** WARNING: IEEEtran.bst: No hyphenation pattern has been}%
\typeout{** loaded for the language `#1'. Using the pattern for}%
\typeout{** the default language instead.}%
\else
\language=\csname l@#1\endcsname
\fi
#2}}
\providecommand{\BIBdecl}{\relax}
\BIBdecl

\bibitem{dnn_face}
Y.~Taigman, M.~Yang, M.~Ranzato, and L.~Wolf, ``{DeepFace: Closing the gap to
  human-level performance in face verification},'' in \emph{Computer Vision and
  Pattern Recognition (CVPR)}, 2014, pp. 1701--08.

\bibitem{dnn_speech}
L.~Deng and et. al, ``{Recent advances in deep learning for speech research at
  Microsoft},'' in \emph{IEEE Intl. Conf. on Acoustics, Speech and Signal
  Processing (ICASSP)}, 2013, pp. 8604--08.

\bibitem{dnn_nlp}
R.~Collobert, J.~Weston, L.~Bottou, M.~Karlen, K.~Kavukcuoglu, and P.~Kuksa,
  ``{Natural language processing (almost) from scratch},'' \emph{J. Machine
  Learning Research}, vol.~12, pp. 2493--2537, 2011.

\bibitem{basu_lora}
V.~M. Suresh and et. al., ``{Powering the IoT through embedded machine learning
  and LoRa},'' in \emph{IEEE World Forum on IoT (WF-IOT)}, 2018, pp. 349--354.

\bibitem{basu_bmi_chapter}
A.~Basu, Y.~Chen, and E.~Yao, ``Big data management in neural implants: The
  neuromorphic approach,'' in \emph{Emerging Technology and Architecture for
  Big-data Analytics}, C.~A. and C.~C.~Y. H, Eds.\hskip 1em plus 0.5em minus
  0.4em\relax Cham: Springer, 2017.

\bibitem{basu_jetcas}
A.~Basu and et. al, ``Low-power, adaptive neuromorphic systems: Recent progress
  and future directions,'' \emph{IEEE Journal on Emerging and Selected Topics
  in Circuits and Systems (JETCAS)}, vol.~8, no.~1, pp. 6--27, 2018.

\bibitem{sarpeshkar_anavsdig}
R.~Sarpeshkar, ``{Analog versus digital: extrapolating from electronics to
  neurobiology},'' \emph{Neural Computation}, vol.~10, no.~7, pp. 1601--38,
  1998.

\bibitem{snn_pfeiffer}
M.~Pfeiffer and T.~Pfeil, ``{Deep learning with spiking neurons: opportunities
  and challenges},'' \emph{Frontiers in Neuroscience}, vol.~12, 2018.

\bibitem{TDNN_bibhu}
B.~D. Sahoo, ``Ring oscillator based sub-1v leaky integrate-and-fire neuron
  circuit,'' in \emph{IEEE Symp. Circuits and Systems (ISCAS)}.\hskip 1em plus
  0.5em minus 0.4em\relax IEEE, 2017, pp. 1--4.

\bibitem{TDNN_ASSCC18}
L.~Everson, M.~Liu, N.~Pande, and C.~H. Kim, ``{A 104.8TOPS/W One-Shot
  Time-Based Neuromorphic Chip Employing Dynamic Threshold Error Correction in
  65nm},'' in \emph{IEEE Asian Solid-State Circuits Conference (ASSCC)}, 2018.

\bibitem{elmpuf_tcas1}
Y.~Chen, Z.~Wang, A.~Patil, and A.~Basu, ``{A 2.86-TOPS/W Current Mirror
  Cross-Bar Based Machine-Learning and Physical Unclonable Function Engine for
  Internet-of-Things Applications},'' \emph{IEEE Trans. on Circuits and
  Systems-I}, vol.~66, no.~6, pp. 2240--52, 2019.

\bibitem{elm_enyi}
E.~Yao and A.~Basu, ``{VLSI Extreme Learning Machine: A Design Space
  Exploration},'' \emph{IEEE Trans. on VLSI}, vol.~25, no.~1, pp. 60--74, 2017.

\bibitem{elm_patil}
A.~Patil, S.~Shen, E.~Yao, and A.~Basu, ``{Hardware Architecture for Large
  Parallel Array of Random Feature Extractors applied to Image Recognition},''
  \emph{Neurocomputing}, vol. 261, pp. 193--203, 2017.

\bibitem{pseudo_RO}
W.~{Bae}, H.~{Ju}, K.~{Park}, S.~{Cho}, and D.~{Jeong}, ``{A 7.6 mW 414 fs
  RMS-jitter 10 GHz phase-locked loop for a 40 Gb/s serial link transmitter
  based on a two-stage ring oscillator in 65 nm CMOS},'' \emph{IEEE J.
  Solid-State Circuits}, vol.~51, no.~10, pp. 2357--2367, 2016.

\bibitem{cicc07}
K.~R. Lakshmikumar and et. al., ``{A Process and Temperature Compensated
  Two-Stage Ring Oscillator},'' in \emph{Custom Integrated Circuits Conference
  (CICC)}, 2007, pp. 691--4.

\bibitem{sengupta2019going}
A.~Sengupta, Y.~Ye, R.~Wang, C.~Liu, and K.~Roy, ``Going deeper in spiking
  neural networks: Vgg and residual architectures,'' \emph{Frontiers in
  neuroscience}, vol.~13, 2019.

\bibitem{rueckauer2017conversion}
B.~Rueckauer, I.-A. Lungu, Y.~Hu, M.~Pfeiffer, and S.-C. Liu, ``Conversion of
  continuous-valued deep networks to efficient event-driven networks for image
  classification,'' \emph{Frontiers in neuroscience}, vol.~11, p. 682, 2017.

\bibitem{arijit_piee}
A.~Raychowdhury, A.~Parihar, and et~.al., ``{Computing With Networks of
  Oscillatory Dynamical Systems},'' \emph{Proc. of IEEE}, vol. 107, no.~1, pp.
  73--89, 2019.

\bibitem{arijit_nature}
S.~Dutta, A.~Parihar, and et~.al., ``{Programmable coupled oscillators for
  synchronized locomotion},'' \emph{Nature Communications}, vol.~10, no. 3299,
  2019.

\bibitem{hajimiri1999jitter}
A.~Hajimiri, S.~Limotyrakis, and T.~H. Lee, ``Jitter and phase noise in ring
  oscillators,'' \emph{IEEE Journal of Solid-state circuits}, vol.~34, no.~6,
  pp. 790--804, 1999.

\bibitem{dudek2000high}
P.~Dudek, S.~Szczepanski, and J.~V. Hatfield, ``A high-resolution cmos
  time-to-digital converter utilizing a vernier delay line,'' \emph{IEEE
  Journal of Solid-State Circuits}, vol.~35, no.~2, pp. 240--247, 2000.

\bibitem{alpha_power_1}
A.~N. T.~Sakurai, ``{Alpha-Power Law MOSFET Model and its Applications to CMOS
  Inverter Delay and Other Formulas},'' \emph{IEEE J. Solid-State Circuits},
  vol.~25, 1990.

\bibitem{alpha_power_2}
K.~A. Bowman, B.~L. Austin, J.~C. Eble, X.~Tang, and J.~D. Meindl, ``A physical
  alpha-power law mosfet model,'' \emph{IEEE Journal of Solid-State Circuits},
  vol.~34, no.~10, pp. 1410--1414, 1999.

\bibitem{Abidi_jitter}
A.~A. Abidi, ``{Phase noise and jitter in CMOS ring oscillators},'' \emph{IEEE
  J. Solid-State Circuits}, vol.~41, no.~8, pp. 1803--1816, 2006.

\bibitem{nature_joshua}
P.~Yao, H.~Wu, and et. al, ``{Fully hardware-implemented memristor
  convolutional neural network},'' \emph{Nature}, vol. 577, 2020.

\bibitem{lee2018integrated}
J.-J. Lee, J.~Park, M.-W. Kwon, S.~Hwang, H.~Kim, and B.-G. Park, ``Integrated
  neuron circuit for implementing neuromorphic system with synaptic device,''
  \emph{Solid-State Electronics}, vol. 140, pp. 34--40, 2018.

\bibitem{basu2010nullcline}
A.~Basu and P.~E. Hasler, ``Nullcline-based design of a silicon neuron,''
  \emph{IEEE Transactions on Circuits and Systems I: Regular Papers}, vol.~57,
  no.~11, pp. 2938--2947, 2010.

\bibitem{leene20180}
L.~B. Leene and T.~G. Constandinou, ``{A 0.006 $mm^2$ 1.2 $\mu$W Analog-to-Time
  Converter for Asynchronous Bio-Sensors},'' \emph{IEEE Journal of Solid-State
  Circuits}, vol.~53, no.~9, pp. 2604--2613, 2018.

\bibitem{li2017174}
S.~Li, A.~Mukherjee, and N.~Sun, ``{A 174.3-dB FoM VCO-Based CT $\Delta\Sigma$
  Modulator With a Fully-Digital Phase Extended Quantizer and Tri-Level
  Resistor DAC in 130-nm CMOS},'' \emph{IEEE Journal of Solid-State Circuits},
  vol.~52, no.~7, pp. 1940--1952, 2017.

\bibitem{young201475db}
B.~Young, K.~Reddy, S.~Rao, A.~Elshazly, T.~Anand, and P.~K. Hanumolu, ``{A
  75dB DR 50MHz BW $3^{rd}$ order CT-$\Delta\Sigma$ modulator using VCO-based
  integrators},'' in \emph{2014 Symposium on VLSI Circuits Digest of Technical
  Papers}.\hskip 1em plus 0.5em minus 0.4em\relax IEEE, 2014, pp. 1--2.

\bibitem{taylor2013reconfigurable}
G.~Taylor and I.~Galton, ``{A reconfigurable mostly-digital delta-sigma ADC
  with a worst-case FOM of 160 dB},'' \emph{IEEE Journal of Solid-State
  Circuits}, vol.~48, no.~4, pp. 983--995, 2013.

\bibitem{kim201411}
Y.~Kim and T.~W. Kim, ``{An 11 b 7 ps resolution two-step time-to-digital
  converter with 3-D Vernier space},'' \emph{IEEE Transactions on Circuits and
  Systems I: Regular Papers}, vol.~61, no.~8, pp. 2326--2336, 2014.

\bibitem{tu2017low}
C.-C. Tu, Y.-K. Wang, and T.-H. Lin, ``{A low-noise area-efficient chopped
  VCO-based CTDSM for sensor applications in 40-nm CMOS},'' \emph{IEEE Journal
  of Solid-State Circuits}, vol.~52, no.~10, pp. 2523--2532, 2017.

\bibitem{jayaraj2019highly}
A.~Jayaraj, M.~Danesh, S.~T. Chandrasekaran, and A.~Sanyal, ``{Highly Digital
  Second-Order $\Delta\Sigma$ VCO ADC},'' \emph{IEEE Transactions on Circuits
  and Systems I: Regular Papers}, vol.~66, no.~7, pp. 2415--2425, 2019.

\bibitem{Jayaraj_ASSCC19}
A.~{Jayaraj}, A.~{Das}, S.~{Arcot}, and A.~{Sanyal}, ``{8.6fJ/step VCO-Based CT
  2nd-Order $\Delta\Sigma$ ADC},'' in \emph{2019 IEEE Asian Solid-State
  Circuits Conference (A-SSCC)}, 2019, pp. 197--200.

\bibitem{Zhong_ASSCC18}
Y.~{Zhong}, S.~{Li}, A.~{Sanyal}, X.~{Tang}, L.~{Shen}, S.~{Wu}, and N.~{Sun},
  ``{A Second-Order Purely VCO-Based CT $\Delta\Sigma$ ADC Using a Modified
  DPLL in 40-nm CMOS},'' in \emph{2018 IEEE Asian Solid-State Circuits
  Conference (A-SSCC)}, 2018, pp. 93--94.

\bibitem{opto-natcomm}
R.~A. John and et. al, ``{Optogenetics inspired transition metal dichalcogenide
  neuristors for in-memory deep recurrent neural networks},'' \emph{Nature
  Communications}, vol.~11, no.~1, pp. 1--9, 2020.

\bibitem{bhaduri_neco}
A.~Bhaduri and et. al., ``{Spiking neural classifier with lumped dendritic
  nonlinearity and binary synapses: a current mode VLSI implementation and
  analysis},'' \emph{Neural Computation}, vol.~30, no.~3, pp. 723--60, 2018.

\bibitem{sroy_tnnls}
S.~Roy and A.~Basu, ``{An online unsupervised structural plasticity algorithm
  for spiking neural networks},'' \emph{IEEE Trans. on Neural Networks and
  Learning Systems}, vol.~28, no.~4, pp. 900--910, 2016.

\bibitem{querlioz}
D.~Querlioz and et. al, ``{Immunity to device variations in a spiking neural
  network with memristive nanodevices},'' \emph{IEEE Trans. on Nanotechnology},
  vol.~12, no.~3, pp. 288--95, 2013.

\bibitem{mnist}
Y.~LeCun, C.~Cortes, and C.~Burges, ``The mnist database of handwritten
  digits,'' \url{http://yann.lecun.com/exdb/mnist/}, 1998.

\bibitem{FMNIST}
H.~Xiao, K.~Rasul, and R.~Vollgraf, ``Fashion-mnist: a novel image dataset for
  benchmarking machine learning algorithms,'' \emph{arXiv preprint
  arXiv:1708.07747}, 2017.

\bibitem{CIFAR10}
A.~Krizhevsky, G.~Hinton \emph{et~al.}, ``Learning multiple layers of features
  from tiny images,'' 2009.

\bibitem{noiseregular}
H.~Noh, T.~You, J.~Mun, and B.~Han, ``{Regularizing deep neural networks by
  noise: Its interpretation and optimization},'' in \emph{Advances in Neural
  Information Processing Systems}, 2017, pp. 5109--5118.

\bibitem{giacomo_neuron}
G.~Indiveri and et. al., ``Neuromorphic silicon neuron circuits,''
  \emph{Frontiers in Neuroscience}, vol.~5, 2011.

\end{thebibliography}

\end{document}